\documentclass[fleqn,usenatbib]{mnras}

\usepackage{newtxtext,newtxmath}

\usepackage[T1]{fontenc}

\DeclareRobustCommand{\VAN}[3]{#2}
\let\VANthebibliography\thebibliography
\def\thebibliography{\DeclareRobustCommand{\VAN}[3]{##3}\VANthebibliography}

\usepackage{graphicx}	
\usepackage{amsmath}	
\usepackage{subcaption}
\usepackage{multirow}
\usepackage{cancel}
\usepackage{soul}
\usepackage{xcolor}
\usepackage[normalem]{ulem}

\title[Variability in $\epsilon$ Ori]{On the origin of variability in $\alpha$ Cygni variable $\epsilon$ Ori (HD 37128) using TESS observations and modelling}

\author[Dasgupta et al.]{Subharthi Dasgupta$^{1}$\thanks{E-mail:522PH1004@nitrkl.ac.in}, Sugyan Parida$^{1}$, Abhay Pratap Yadav$^{1}$\thanks{E-mail: yadavap@nitrkl.ac.in}, Wolfgang Glatzel$^{2}$ and Michaela Kraus$^{3}$
\\
$^{1}$Department of Physics and Astronomy, National Institute of Technology Rourkela - 769008, Odisha, India\\
$^{2}$ Institut f\"ur Astrophysik und Geophysik, Georg-August-Universit\"at G\"ottingen, Friedrich-Hund-Platz 1, D-37077 G\"ottingen, Germany
\\
$^{3}$ Astronomical Institute, Czech Academy of Sciences, Fri\v{c}ova 298, CZ-251 65 Ond\v{r}ejov, Czech Republic
}

\date{Accepted XXX. Received YYY; in original form ZZZ}

\pubyear{\the\year{}}

\begin{document}
\label{firstpage}
\pagerange{\pageref{firstpage}--\pageref{lastpage}}
\maketitle

\begin{abstract}

$\epsilon$ Ori (HD 37128) is an  $\alpha$ Cygni variable characterized by irregular and small amplitude variations. 
From TESS observations, we find the presence of stochastic low-frequency variability in this star. We have constructed a sequence of models for this star in the mass range of 30 to 70 M$_{\odot}$, using recently derived values of luminosity (log $(L/L_{\odot})$ = 5.92) and effective temperature. In these considered models, both radial and non-radial linear stability analyses have been performed. Low-order radial modes are excited in models having mass below 62 M$_{\odot}$. These radially excited modes have periods ranging from 6.8 days for the fundamental mode to a few hours for higher-order modes. Similar to the case of radial modes, several non-radial modes are found to be unstable in models having higher luminosity-to-mass ratios. Linear stability analysis for the case of $l$ = 2 and $l$ = 4 reveals the presence of a strongly unstable mode in models having a mass below 40 M$_{\odot}$. This mode is found to be unstable in all the considered models and the strength of the instability varies as a function of harmonic degree. The non-adiabatic reversible approximation reveals that the origin of instabilities associated with the low-order modes is indeed linked with strange modes. To find out the consequence of radial instabilities, non-linear numerical simulations have been performed in selected models of $\epsilon$ Ori. In the non-linear regime, these instabilities lead to the envelope inflation, finite amplitude regular and irregular pulsations consistent with an $\alpha$ Cygni variable.

\end{abstract}




\section{Introduction} \label{introduction}

$\alpha$ Cygni variables are supergiant stars of spectral class B and A which display semi-regular variability in both the photometric and spectroscopic domains \citep{sterken1977}. Such variability has been observed in the luminosity and radial velocity of supergiants by \citet{abt1957}. For the prototype of this class $\alpha$ Cyg (Deneb), \citet{Lucy_1976} extensively studied the radial velocity data and suggested that the quasi-periodic variability could be due to the excitation of many pulsation modes in the star. Further photometric studies of this kind of stars found that they exhibited microvariations in the luminosity which were semi-regular or erratic in nature \citep{van1989, van1998, lefevre2009}. The $\alpha$ Cygni variables are also characterized by significant variability in the wind and photospheric lines \citep{kaufer1996, kaufer1997}. Multiperiodic variability seems to be present and while strict periodicities are not found, quasi-periods ranging from few days to months were observed. Their origin is not yet fully understood.   

Theoretical models of massive, luminous stars are well known to exhibit instabilities, including so-called 'strange mode' instabilities \citep{kiriakidis1993,saio2011}. Particularly, in the context of $\alpha$ Cygni variables, \citet{gautschy1992} found non-radial strange mode instabilities in models resembling $\alpha$ Cyg for high luminosity-to-mass ratios. But their excitation required a far too lower mass to be realistic. A further attempt at studying the instabilities in Deneb was done by \citet{gautschy2009} who showed that non-radial modes with periods of few days could be excited by the presence of a convection zone above the hydrogen burning shell in models close to the position of Deneb on the Hertzsprung-Russell (HR) diagram. \citet{saio2013} studied the pulsational instabilities in evolutionary models of blue supergiants and found that most radial and non-radial modes (including strange modes) are excited in models on the blue loop after the red supergiant stage. These modes have periods which appear to be consistent with the observed periods in several $\alpha$ Cygni variables. As such, the authors identified the $\alpha$ Cygni variables as He-burning stars which are evolving on the blue loop. The surface abundances (He, C, N, O) predicted by their models, however, are much higher than those observed in $\alpha$ Cygni stars. 
\citet{georgy2014} showed that the use of the Ledoux criterion for the onset of convection can significantly improve the agreement of the observed surface abundances with models.

$\epsilon$ Orionis (HD 37128) is an $\alpha$ Cygni variable which is the central star of Orion's belt. It is a blue supergiant classified as B0Ia \citep{lesh1968}. This star has been extensively studied and several determinations of its fundamental parameters are available in the literature \citep{auer1972, lamers1974, mcerlean1998, kudritzki1999, crowther2006, searle2008}. A multiwavelength study of $\epsilon$ Ori in the optical, UV and X-ray domain was performed by \citet{puebla2016}. The analysis was based on optical data from the \textit{POLARBASE} archive, UV data from \textit{IUE}, \textit{HST} and the \textit{COPERNICUS} satellite, and X-ray data from \textit{Chandra} and \textit{XMM-Newton}. The derived fundamental parameters such as the effective temperature T$_\mathrm{eff}$ = 27000$\pm$500 K and a surface gravity of $\log$ g = 3.0$\pm$0.05, were found to be consistent with previous values. Two distance estimates for the star from the \textit{HIPPARCOS} catalogue were used: 412 pc \citep{perryman1997} and a more recently determined value of 606 pc \citep{vanleeuwen2007}. 
For the lower distance (412 pc), the luminosity was estimated to be $\log (L_*/L_{\odot})$ = 5.60 whereas for the larger value of distance (606 pc), a luminosity value of $\log (L_*/L_{\odot})$ = 5.92 was obtained, which is the highest value derived for this star.
The CNO abundances as derived from X-ray, UV and optical analyses are consistent and show slight enhancement in nitrogen and depletion of carbon and oxygen \citep{puebla2016}.

$\epsilon$ Ori (HD 37128) exhibits photometric as well as spectroscopic variabilities. The variability of this star has been most extensively studied in the optical domain, particularly using the H$\alpha$ line profile. A study of the H$\alpha$ profile by \citet{ebbets1982} found variabilities of the order of 1 to 10 days. Subsequently, \citet{morel2004} found significant variability in the H$\alpha$ and HeI $\uplambda 6678$ lines. Variations in the H$\alpha$ profile were found on the timescale of hours to days, and two periods of 0.781 d and 18.2 d were found in the H$\upalpha$ data. 
The period of 18.2 d was of the order of the maximum rotational period of the star (17.6 d). The maximum value of the rotational period was estimated by \citet{morel2004} using the value of $v \sin i$ ($\sim$ 91 kms$^{-1}$) determined by \citet{howarth1997} and assuming an equator-on view of the star. Any uncertainty in the inclination angle and  rotation velocity can considerably affect the value of the rotation period. In a recent study, \citet{deburgos2024} derived a $v \sin i$ value of 52 kms$^{-1}$ for this star which implies a longer rotation period ($>$ 30 d).
\citet{prinja2004} conducted a time series analysis of both H$\upalpha$ as well as several photospheric lines. Three significant periods of 1.9, 6.6 and 9.7 d were consistently identified in H$\upalpha$, H$\upbeta$ and He I $\uplambda$6678 lines, although their origin could not be confirmed. An extensive study of the optical line profile variability of $\epsilon$ Ori was carried out by \citet{thompson2013}. Their analysis revealed several periods in the variability of the order of 2-8 days, with no direct connection to the rotational period. Some of these periods were found both in the H$\alpha$ and HeI lines, indicating a possible wind-photosphere connection. These also exhibit modulation in their phase diagrams, and their origin could be due to stellar pulsation \citep{thompson2013}. \citet{martins2015} investigated the line profile variability of seven OB stars including $\epsilon$ Ori and found significant variability in its  wind lines. However, several photospheric lines did not show any clear changes. From the H$\alpha$ line variability, they found a period of about two days, which was overall consistent with the results of \citet{prinja2004} and \citet{thompson2013}. However, they claimed that their sampling was too coarse for a proper period identification. The cause of the spectroscopic variability of the star is not fully understood.

Recently, the star has been observed by the Transiting Exoplanet Survey Satellite (TESS; \citealt{Ricker_2015}). \citet{burssens2020} studied the one-sector TESS data and reported mainly stochastic low frequency variability. As the dominant frequency falls in the range of the rotational modulation, they suggested that it could be associated with rotational effects. This star was included in a sample of O and B-type stars studied by \citet{bowman2020} who characterized their stochastic low frequency variability and found strong evidence of internal gravity waves being excited in these massive stars. In this paper, we aim to study the radial and non-radial instabilities in models of $\epsilon$ Ori together with recent space-based photometric observations to understand the origin of observed variabilities.

Our analysis of the TESS data from two sectors (6 and 32) for this star have been presented in Section \ref{TESS}. 
In order to explore whether pulsation could be related to some of the observed variability of the star, we have constructed stellar models with the observed parameters of HD 37128 and performed non-adiabatic linear stability analysis followed by non-linear simulations. The stellar models adopted for our study are described in Section \ref{models}. The results of the non-adiabatic linear stability analysis for radial and non-radial perturbations are shown in Section \ref{lna}. Outcomes of the non-linear simulations are described in Section \ref{nonlinear}. The results are discussed in Section \ref{discussion} and our conclusions follow in Section \ref{conclusions}.

\section{TESS photometry} \label{TESS}

\begin{figure*}
    \centering
    \begin{subfigure}{0.45\textwidth}
        \centering
        \includegraphics[width=\textwidth]{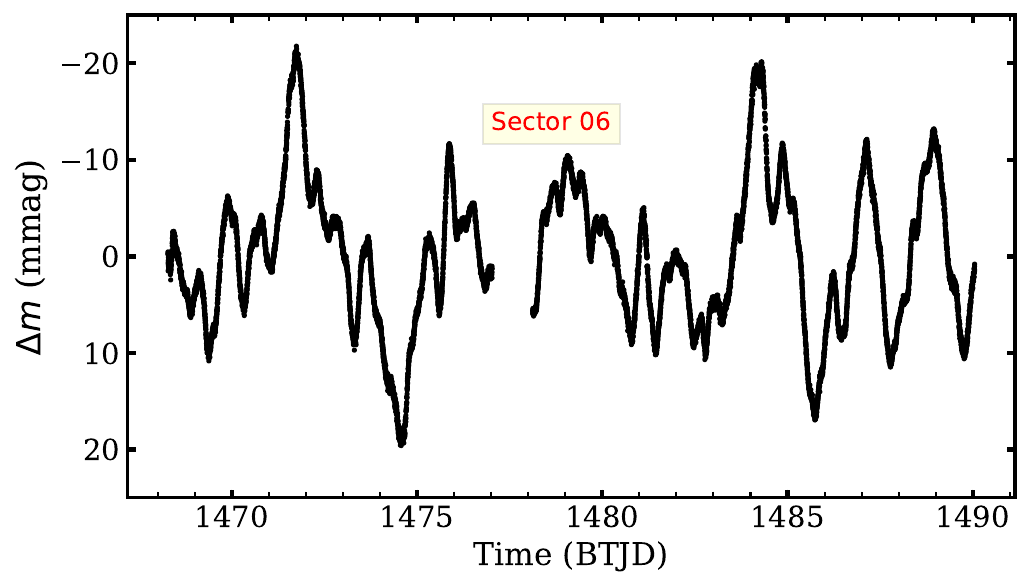}
        \caption{Lightcurve Sector 06}
        \label{fig1:sub1}
    \end{subfigure}
    \hspace{0.05\textwidth} 
    \begin{subfigure}{0.45\textwidth}
        \centering
        \includegraphics[width=\textwidth]{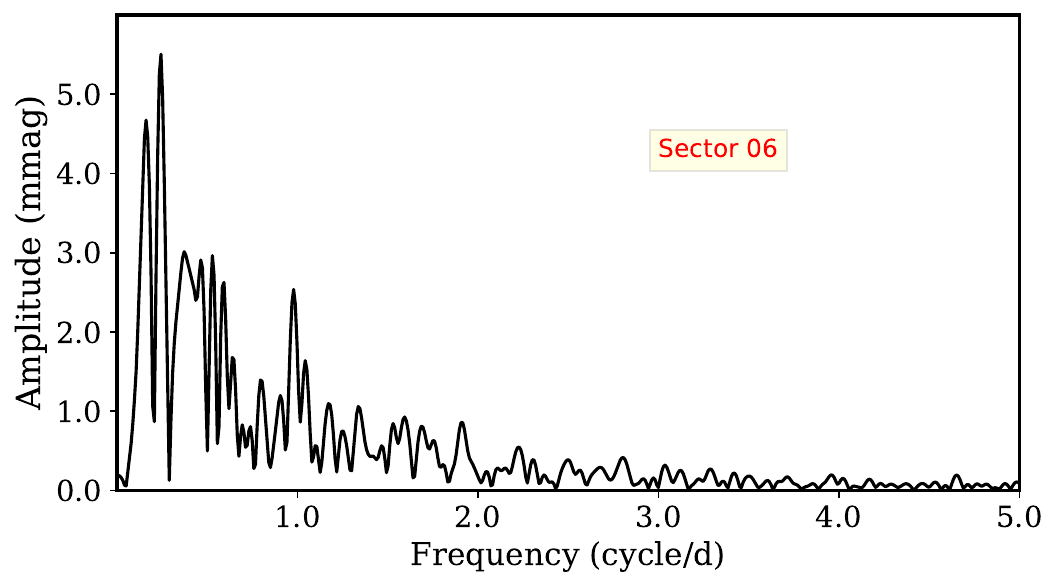}
        \caption{Periodogram Sector 06}
        \label{fig1:sub2}
    \end{subfigure}
    
    \vspace{0.5cm} 
    
    \begin{subfigure}{0.45\textwidth}
        \centering
        \includegraphics[width=\textwidth]{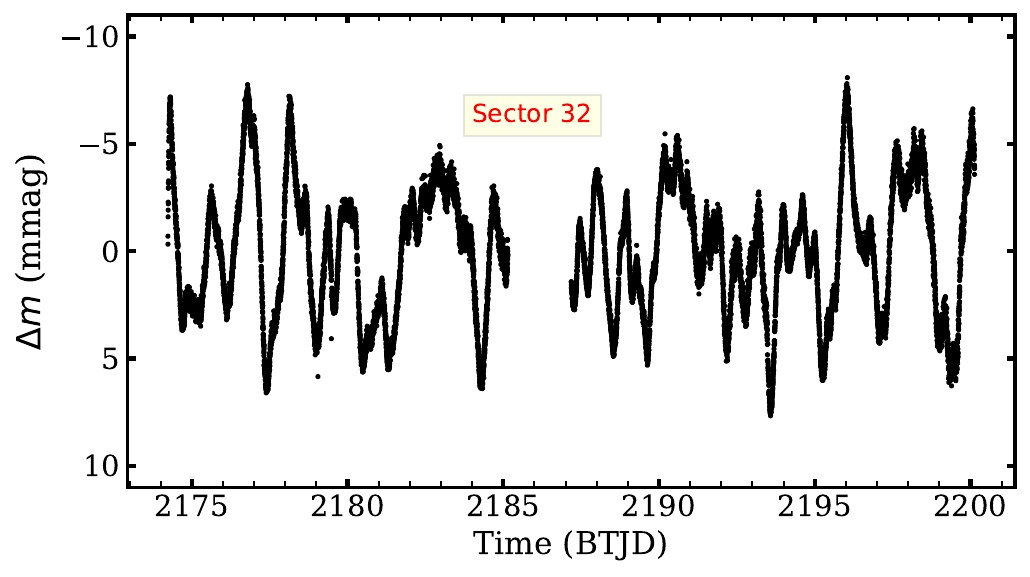}
        \caption{Lightcurve Sector 32}
        \label{fig1:sub3}
    \end{subfigure}
    \hspace{0.05\textwidth} 
    \begin{subfigure}{0.45\textwidth}
        \centering
        \includegraphics[width=\textwidth]{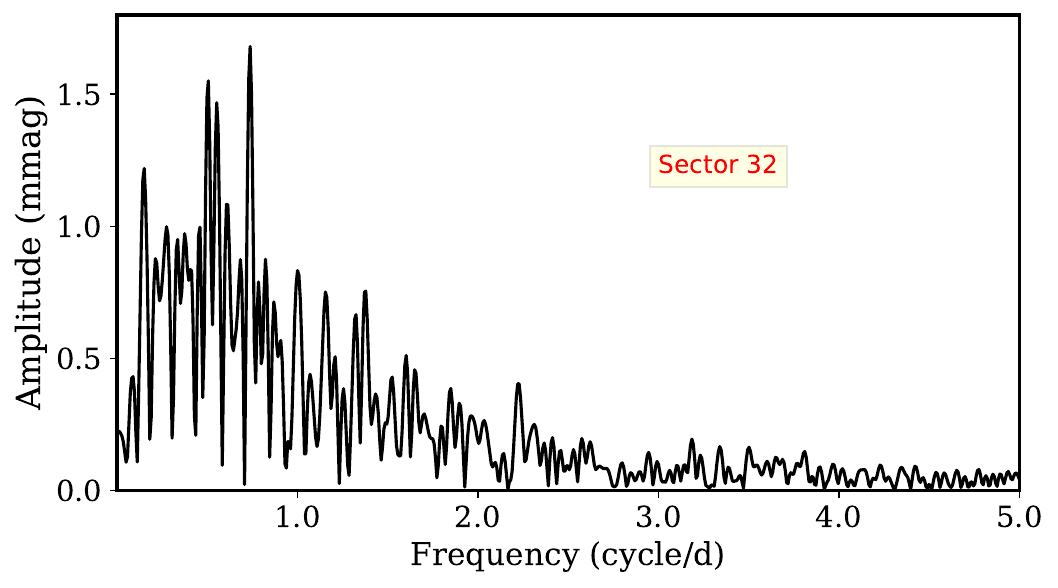}
        \caption{Periodogram Sector 32}
        \label{fig1:sub4}
    \end{subfigure}

     \vspace{0.5cm} 

    \caption{The left panels (\ref{fig1:sub1} and \ref{fig1:sub3}) show the two-minute cadence TESS light curves for Sectors 6 and 32, extracted using the SPOC pipeline apertures, while the right panels (\ref{fig1:sub2} and \ref{fig1:sub4}) display the corresponding amplitude spectra of $\epsilon$~Ori.}

    \label{fig1}
\end{figure*}

To date, HD\,37128 (TIC 427451176) has been observed by the TESS \citep{Ricker_2015} in two sectors: Sector 6 (2018 December 11–2019 January 7) and Sector 32 (2020 November 19–December 17). Each sector provides Full Frame Images (FFIs) at different cadences—30 minutes in Sector 6 (primary mission) and 10 minutes in Sector 32 (first extension). Additionally, HD\,37128 was observed as a high-cadence target with a 2-minute sampling in both sectors. All essential data reduction steps—including bias subtraction, flat-fielding, background subtraction, and pixel-level calibration—are performed by the Science Processing Operations Center (SPOC) pipeline \citep{Jenkins_2016}. In this work, we use the 2-minute high-cadence data, for which SPOC pipeline extracts photometry using an optimal aperture. The resulting time series are corrected for crowding, instrumental systematics, and the finite aperture size, and are made available through the Mikulski Archive for Space Telescopes (MAST). These data were accessed using the Python package \texttt{Lightkurve} \citep{2018ascl.soft12013L}.
The light curves of HD\,37128 from Sectors 6 and 32 were analysed separately, accounting for small mean-flux differences and the two-year gap between observations. The fluxes were converted to the magnitude scale using $m = -2.5 \log_{10}(\mathrm{flux})$, after which the mean magnitude was subtracted. To study the time series in the frequency domain, Lomb-Scargle periodograms \citep{Lomb_1976,Scargle_1982,VanderPlas_2018} were computed, as they are well suited for unevenly spaced data. The resulting light curves exhibit amplitude scatters of roughly 40 mmag in Sector 06 and about 15 mmag in Sector 32. The brightness  in millimagnitudes as a function of Barycentric TESS Julian Date (BTJD) are displayed in the left panels of Fig.~\ref{fig1} for Sectors 06 and 32. The corresponding periodograms, shown as a function of frequency (cycle/d), are presented in the right panels of Fig.~\ref{fig1}.

A standard pre-whitening procedure was applied to search for periodicities in the data \citep{burssens2020, bowman2020}.  Frequencies were searched up to the Nyquist frequency (360 d\(^{-1}\)) and, given the short observational baseline of each sector, any frequencies below the formal resolution were disregarded. The significance of the peak is assessed by taking into account the presence of stochastic low frequency variability. No significant (S/N $> 5$) peaks were detected in either sector. In Sector 6, the dominant peak at $0.2415$ $\mathrm{d}^{-1}$ is marginal ($3 < \mathrm{S/N} < 5$) and may be related to rotational effects, as it lies within the rotational modulation frequency range as estimated in \cite{burssens2020}. In Sector 32, the dominant peak at $0.7366$ $\mathrm{d}^{-1}$ is below marginal. The dominant frequencies in each sector, along with their amplitudes, corresponding errors, and S/N, are listed in Table \ref{tab1}. This indicates that the photometric variability of $\epsilon$~Ori is not dominated by long-lived, coherent peaks in the TESS bandpass.

\begin{table}
\centering
\caption{Dominant frequencies in each sector, along with their amplitudes, frequency errors, amplitude errors, and S/N.}
\label{tab1}
\begin{tabular}{lcccccc}
\hline
Sector & Frequency & $\sigma_{\rm freq}$ & Amplitude & $\sigma_{\rm amp}$ & S/N \\
       & (d$^{-1}$) & (d$^{-1}$) & (mmag) & (mmag) &  \\
\hline
6  & 0.2415 & 0.0025 & 5.6400 & 0.5562 & 3.0179 \\
32 & 0.7366 & 0.0019 & 1.6809 & 0.1539 & 2.9695 \\
\hline
\end{tabular}
\end{table}

\begin{figure*}
    \centering
    \begin{subfigure}{0.45\textwidth}
        \centering
        \includegraphics[width=\textwidth]{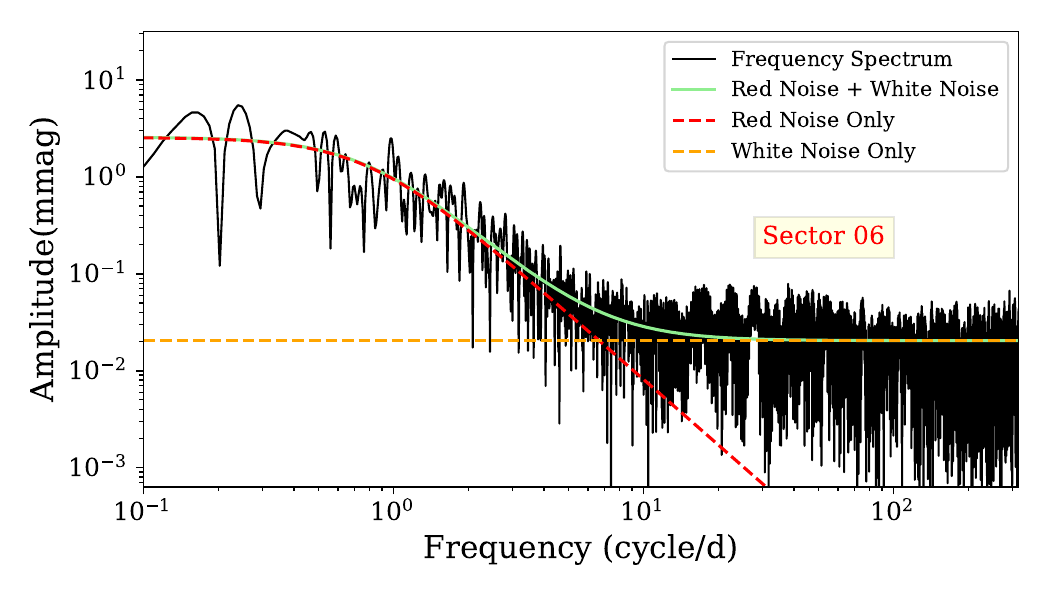}
        \caption{Periodogram Sector 06}
        \label{fig2:sub1}
    \end{subfigure}
    \hspace{0.05\textwidth} 
    \begin{subfigure}{0.45\textwidth}
        \centering
        \includegraphics[width=\textwidth]{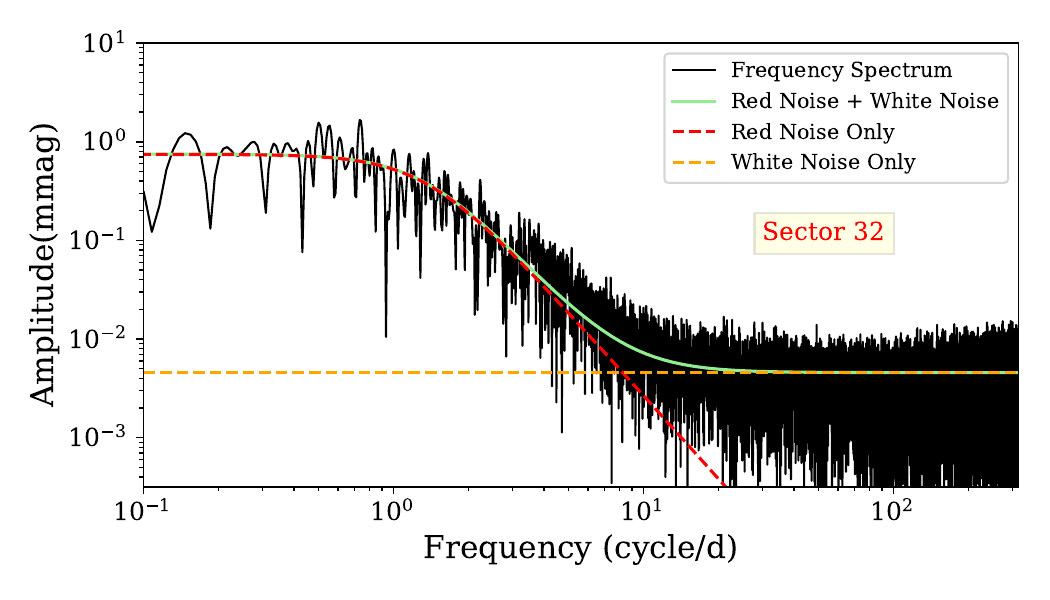}
        \caption{Periodogram Sector 32}
        \label{fig2:sub2}
    \end{subfigure}

    \caption{The left and right panels (\ref{fig2:sub1} and \ref{fig2:sub2}) show the best-fitting semi-Lorentzians for the power spectrum for the light curves in Fig.~\ref{fig1} for Sectors 6 and 32 respectively, based on the parameters in Table 2. The black line is the power spectrum, while the green line represents the best-fit red + white noise model. These two are also plotted separately (red and yellow, respectively).}
    \label{fig2}
\end{figure*}

\subsection{Stochastic low frequency variability} \label{tt}
Given the absence of any significant coherent peaks, we focus on characterising the stochastic low-frequency background present in the TESS photometry.
The periodograms of both sectors show a gradual increase in power towards low frequencies, commonly referred to as red noise. Such stochastic low-frequency variability is ubiquitous in massive stars and has been widely reported in the literature \citep{Blomme_2011,Rauw_2019,Bowman_2019b}.
Following previous studies \citep{Bowman_2019b,bowman2020}, we characterise this variability by fitting a semi-Lorentzian function with an additional white-noise component:
\begin{equation}
\alpha(\nu)=\frac{\alpha_0}{1+\left(\frac{\nu}{\nu_{\mathrm{char}}}\right)^\gamma}+\alpha_w .
\end{equation}
Here, $\nu$ is the frequency, $\alpha_0$ represents the red-noise amplitude at zero frequency, $\gamma$ is the slope of the red-noise component, $\nu_{\mathrm{char}}$ is the characteristic frequency whose inverse corresponds to the typical timescale (or mean lifetime) of the physical structures responsible for the stochastic variability, and $\alpha_w$ denotes the frequency-independent white-noise level.

\begin{figure}
    \centering
    \includegraphics[width=\columnwidth]{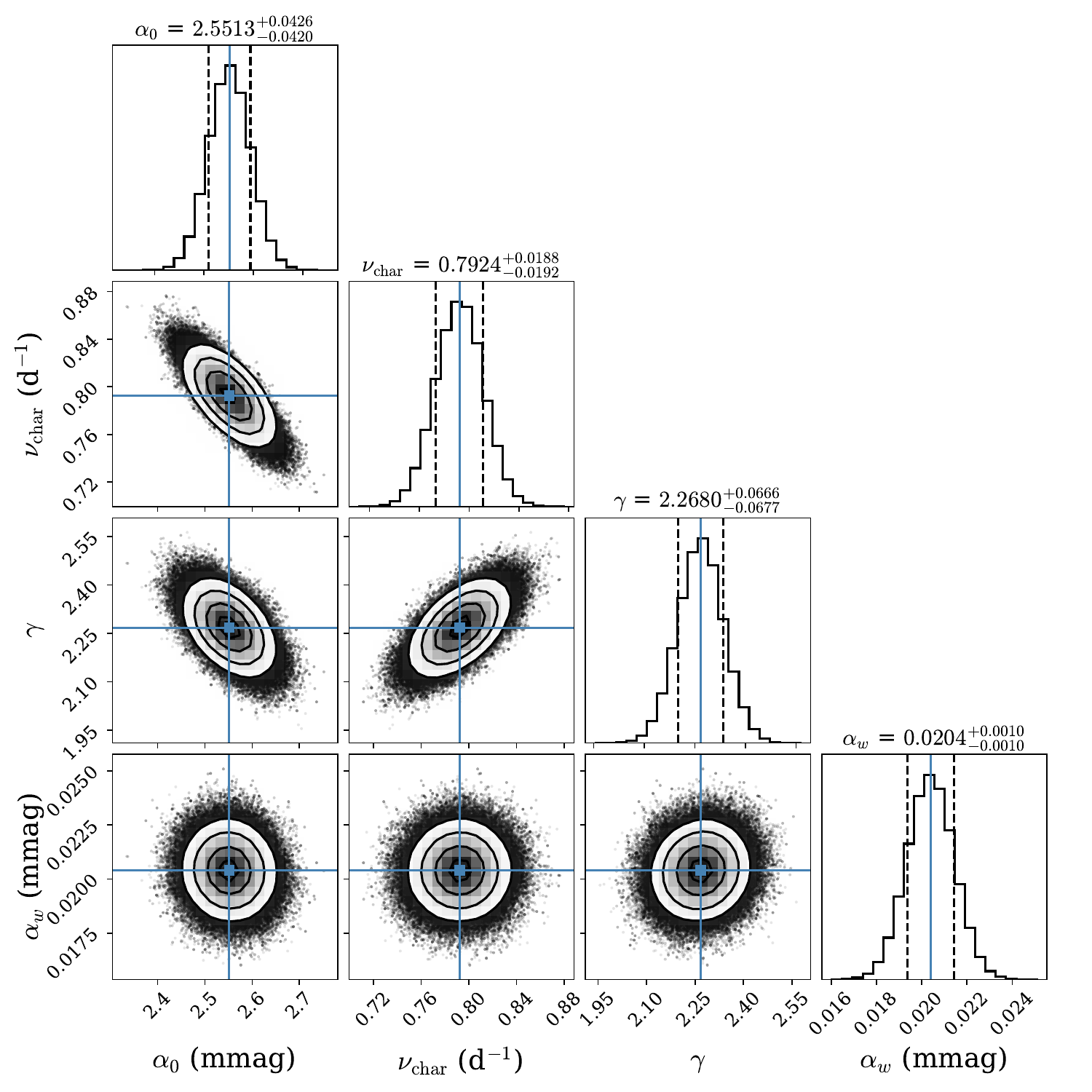}  
    \caption{The posterior distributions for the red-noise parameters of $\epsilon$ Ori in Sector 6 are shown, with dashed vertical lines indicating the 16th, 50th, and 84th percentiles (from left to right). A strong negative correlation is observed between ($\alpha_0$, $\nu_{\mathrm{char}}$) and ($\alpha_0$, $\gamma$), while ($\nu_{\mathrm{char}}$, $\gamma$) exhibits a strong positive correlation.}
    \label{fig3}
\end{figure}

\begin{figure}
    \centering
    \includegraphics[width=\columnwidth]{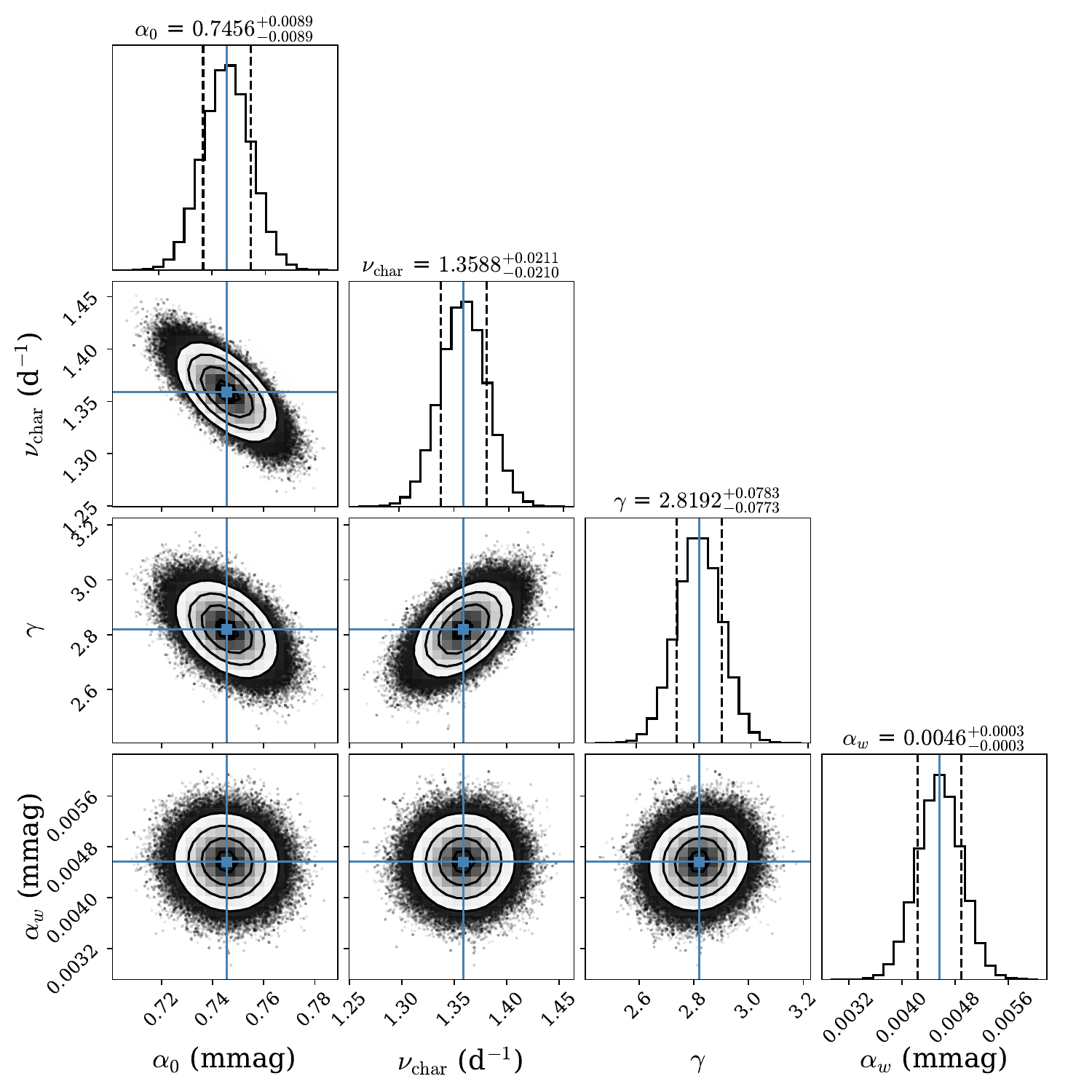}  
    \caption{The posterior distributions for the red-noise parameters of $\epsilon$ Ori in Sector 32 are shown, with dashed vertical lines indicating the 16th, 50th, and 84th percentiles (from left to right). A strong negative correlation is observed between ($\alpha_0$, $\nu_{\mathrm{char}}$) and ($\alpha_0$, $\gamma$), while ($\nu_{\mathrm{char}}$, $\gamma$) exhibits a strong positive correlation).}
    \label{fig4}
\end{figure}

We fit the amplitude spectrum using a custom Markov chain Monte Carlo (MCMC) method using the python EMCEE package \citep{emcee}. Fig.~\ref{fig2} shows the fit of the amplitude spectrum for both sectors 6 and 32. Posterior distribution for the parameters for sector 6 and 32 are shown in the Fig.~\ref{fig3} and Fig.~\ref{fig4} respectively. The best fit parameters along with their uncertainties is mentioned in Table \ref{tab2}. The fit values are found to be well within the uncertainties when compared to previous studies of this star \citep{bowman2020}. We note an order-of-magnitude decrease in the values of $\alpha_0$ and $\alpha_w$ between the sectors . In contrast, the increase in $\gamma$ and $\nu$ is particularly interesting, as these parameters may be related to the intrinsic properties of the star. Several mechanisms have been proposed to explain the stochastic low frequency variability. These include sub-surface convection \citep{cantiello2009, cantiello2021}, stellar winds \citep{krticka2021} or internal gravity waves excited by core convection \citep{aerts2015, bowman2020, thompson2024}.

\begin{table}
\centering
\caption{Optimised parameters for the morphology of low frequency variability using a Bayesian MCMC fitting method.}
\label{tab2}
\setlength{\tabcolsep}{3pt}  
\resizebox{\columnwidth}{!}{
\begin{tabular}{lcccccc}
\hline
Sector & $\alpha_0$ & $\nu_{char}$ & $\gamma$ & $\alpha_w$ &\\
       & (mmag) & (d$^{-1}$) &  & (mmag) &  \\
\hline
6  & 2.5525$^{+0.0420}_{-0.0422}$ & 0.7918$^{+0.0190}_{-0.0189}$ & 2.2660$^{+0.0674}_{-0.0671}$ & 0.0204$^{+0.0010}_{-0.0010}$ &  \\[3pt]
32 & 0.7458$^{+0.0089}_{-0.0089}$ & 1.3585$^{+0.0212}_{-0.0213}$ & 2.8175$^{+0.0782}_{-0.0772}$ & 0.0046$^{0.0003}_{-0.0003}$ &   \\
\hline
\end{tabular}
}
\end{table}

\begin{figure}
    \centering
    \includegraphics[width=0.95\columnwidth]{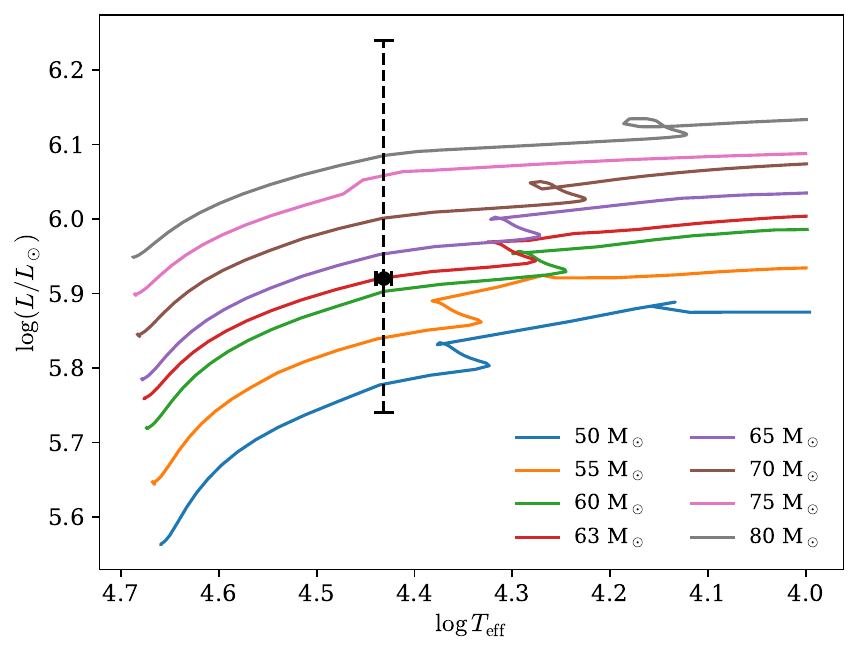}  
    \caption{Evolutionary tracks of models with initial masses of 50, 55, 60, 63, 65, 70, 75 and 80 M$_\odot$. The position of HD 37128 according to its observed effective temperature and luminosity is marked by a black dot. The dashed lines indicate the errors in the observed luminosity and effective temperature \citep{puebla2016}. The high uncertainty in luminosity is due to the large errors in distance.}
    \label{fig5}
\end{figure}

\section{Models for $\epsilon$ Ori (HD 37128)} \label{models}

In order to construct models of $\epsilon$ Ori, the most recent values of the effective temperature (T$_\mathrm{eff}$ = 27000 K) and luminosity $\log (L_*/L_{\odot})$ = 5.92 as determined by \citet{puebla2016} on the basis of the new \textit{HIPPARCOS} distance (606 pc) are considered. 
Evolutionary tracks for non-rotating models having initial masses of 50, 55, 60, 63 and 65 M$_\odot$ respectively are shown in Fig.~\ref{fig5} where the dot corresponds to the observed parameters of HD 37128 in the Hertzsprung-Russell (HR) diagram. The evolutionary tracks have been generated using the MESA \citep{Paxton2011, Paxton2013, Paxton2015,Paxton2018,Paxton2019, Jermyn2023} code using solar chemical composition and a mixing length parameter of 1.5. Also, the Vink mass loss scheme is adopted for the hot wind regime. There is significant uncertainty in the mass loss rates for massive stars due to wind inhomogeneities. 
In recent studies \citep{sundqvist2019,bjorklund2021,hawcroft2021}, mass-loss rates have been observed to be 2 to 3 times lower than predicted by the mass-loss scheme \citep{vink2001}.  
Accordingly, we use a scaling factor of 0.4 \citep[see e.g.,][]{Groth2020} for the Vink mass-loss scheme.
The model which matches the temperature and luminosity of the star corresponds to a track for a model of initial mass 63 M$_\odot$, with the corresponding mass of the model at the location of the star being close to 56.3 M$_\odot$.

Due to large uncertainties in the determination of distance, there is also a significant uncertainty in the luminosity and hence in the determination of mass. Therefore, for linear stability analysis, we have considered models of $\epsilon$ Ori having mass ranging from 30 M$_\odot$ to 70 M$_\odot$. In the present study, we have taken the solar chemical composition (Z = 0.02, Y = 0.28, X = 0.70) for constructing these models. For the given luminosity, effective temperature, chemical composition and mass, models are constructed by integrating the equations of stellar structure-mass conservation, hydrostatic equilibrium and energy transport- from the surface to a point in the interior having a temperature of the order of 10$^7$K. At the photosphere, the Stefan-Boltzmann Law and photospheric pressure are used as initial conditions. 
Rotation and magnetic fields are disregarded. The OPAL opacity tables \citep{rogersiglesias1992,iglesias1996,rogersswenson1996} are used for the computation. The onset of convection is defined by Schwarzschild's criterion and standard mixing length theory with mixing length parameter $\upalpha$=1.5 \citep{bohmvitense1958} has been used.

\section{Linear Stability Analysis} \label{lna}

\subsection{Radial analysis}
\begin{figure*}
    \centering
    \begin{subfigure}{0.45\textwidth}
        \centering 
        \includegraphics[width=\textwidth]{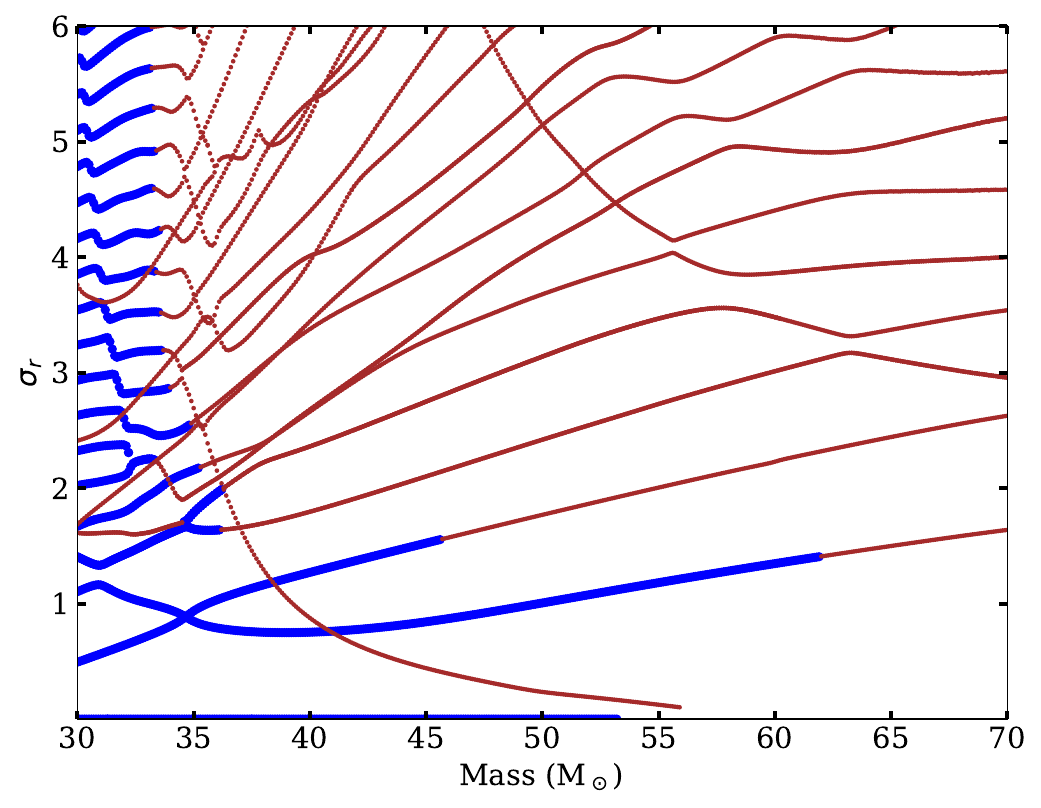}
        \caption{}
        \label{fig6:sub1}
    \end{subfigure}
    \hfill
    \begin{subfigure}{0.45\textwidth}
        \centering
        \includegraphics[width=\textwidth]{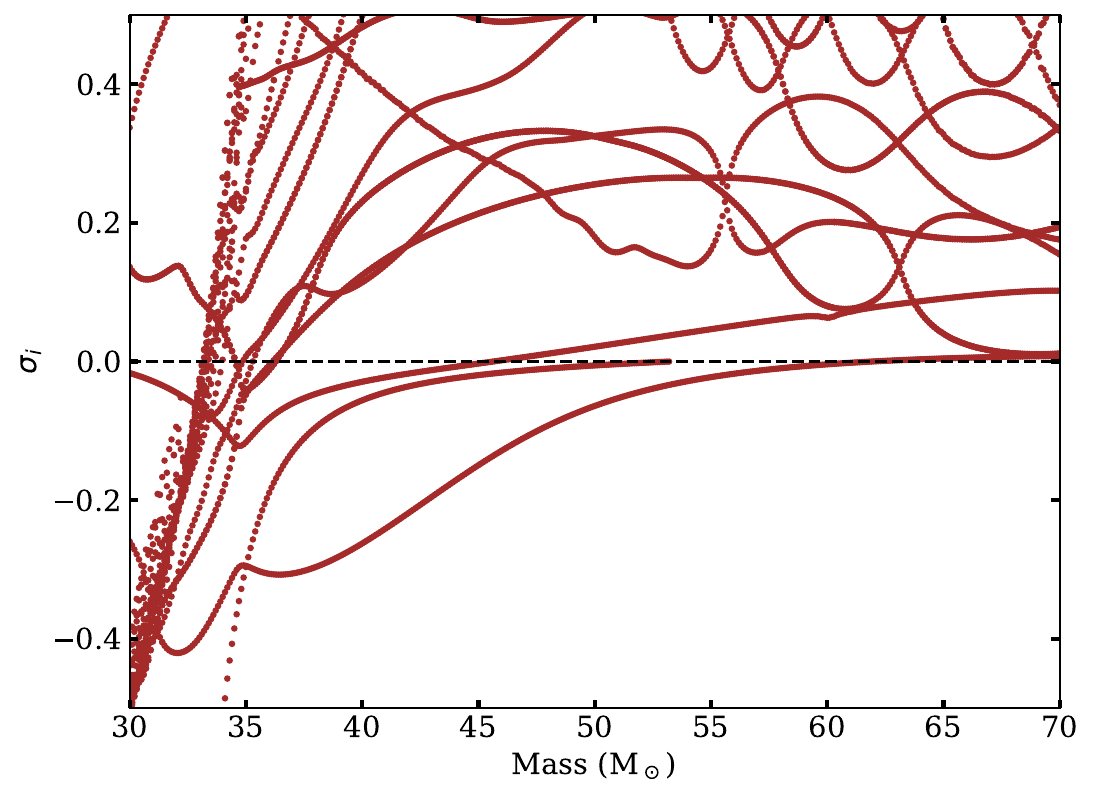}
        \caption{}
        \label{fig6:sub2}
    \end{subfigure}
    \caption{Real (left) and imaginary (right) parts of eigenfrequencies plotted as a function of mass for models of HD 37128 ranging from 30 to 70 M$_\odot$ with effective temperature T$_\mathrm{eff}$ = 27000 K, luminosity $\log (L/L_\odot)$ = 5.92 and a solar chemical composition (Z = 0.02). Thick blue lines in the real part indicate excited modes, which correspond to a negative imaginary part.} 
    \label{fig6}
\end{figure*}

We have performed a linear stability analysis with respect to radial perturbations for the models of $\epsilon$ Ori having the parameters described in Section \ref{models}. The analysis is performed on the basis of the linearized pulsation equations as given by \citet{gg1990b}. The set of pulsation equations together with four boundary conditions (two at the surface and two at the bottom of the stellar envelope) forms a fourth-order boundary eigenvalue problem. This is solved using the Riccati method which has been successfully applied to stellar pulsation problems by \citet{gg1990a}. In this method, the linear boundary value problem is converted into a non-linear initial value problem with unambiguous initial conditions. The complex eigenfrequency $\sigma$ is the only free parameter of the problem. The eigenvalues can be determined by simply finding the roots of a complex determinant function. Initial guesses for the zeros of the determinant function can be obtained by observing the variation of this function in the complex plane. The location of an eigenvalue is indicated by the local minima along the run of the determinant function and the value at that point can be used as an initial guess for further iteration to find the exact eigenfrequencies. It should be noted that the initial guesses do not depend on any approximation of the problem (such as the adiabatic approximation) but rather consider the entire set of perturbation equations. This makes the Riccati method particularly suitable for highly non-adiabatic pulsations in which the non-adiabatic eigenfrequencies may differ significantly from the adiabatic solutions. Also, since the method does not suffer from the problem of unknown initial conditions, it is numerically stable and can be used to compute eigenfrequencies of high order modes  with any prescribed accuracy.

\begin{figure}
    \centering
    \includegraphics[width=0.95\columnwidth]{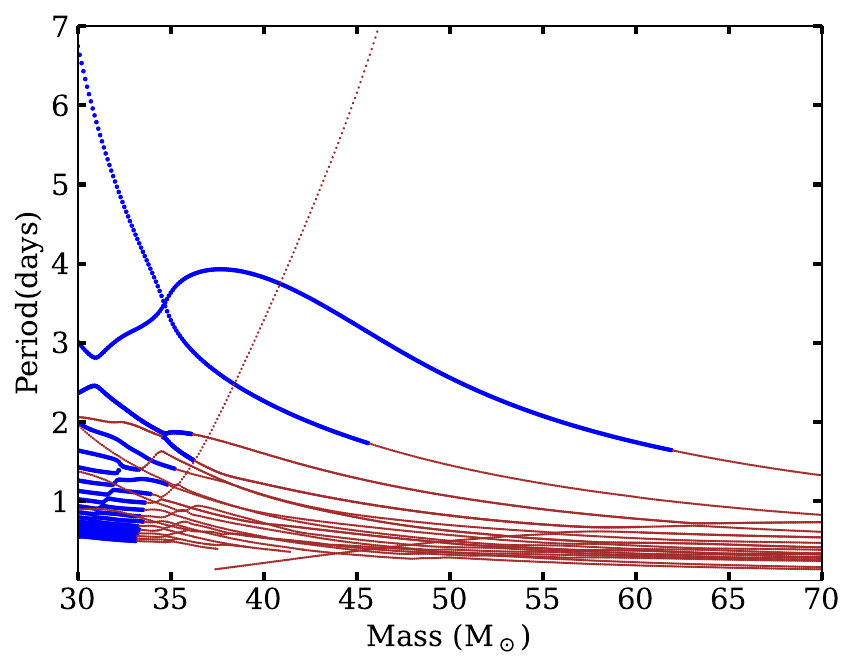}  
    \caption{Period associated with different modes is plotted as a function of mass for models of HD 37128 having solar chemical composition (Z = 0.02). Excited modes are indicated by thick blue lines.}
    \label{fig7}
\end{figure}

We obtain complex eigenfrequencies ($\upsigma$), which are normalized by the global free fall time scale ($\sqrt{R^3/3GM}$, where $R$ is the stellar radius, $G$ the gravitational constant and $M$ denotes the stellar mass) of the corresponding models. The time dependence of the perturbations is taken to be of the form $e^{i\sigma t}$, where the real part of the eigenfrequency ($\upsigma_r$) correspond to the pulsation frequencies while the imaginary part indicates excitation ($\upsigma_i$<0) or damping ($\upsigma_i$>0). Since the interaction of pulsation and convection is not yet well understood, the 'frozen in approximation' introduced by \citet{baker1965} is used for the present analysis, in which the Lagrangian perturbation of the convective flux is assumed to vanish. 

The results of the linear stability analysis are represented in the form of a so-called 'modal diagram'\citep{wood1976,gg1990a,saio1998} where both the real and imaginary parts of the eigenfrequencies are plotted as a function of some fundamental stellar parameter such as stellar mass or effective temperature. Fig.~\ref{fig6} shows a modal diagram where the eigenfrequencies are plotted as a function of mass (with a fixed luminosity and temperature) for the corresponding models-real part in Fig.~\ref{fig6:sub1} (left) and imaginary in Fig.~\ref{fig6:sub2} (right). The thick blue lines indicate those eigenfrequencies whose imaginary parts are negative and thus are unstable modes. From Fig.~\ref{fig6} it can be seen that large number of modes are unstable for models with masses less than 35 M$_\odot$ and no unstable modes are present in models having masses greater than 62 M$_\odot$. For a model of 35 M$_\odot$, real part of the eigenfrequency of two lower order unstable modes are showing interaction. A monotonically unstable ($\sigma_r=0$) mode \citep{saio2011,yadav2016} is found for models below 55 M$_\odot$. The strength of this instability grows steeply for models below 35$M_\odot$. From Fig.~\ref{fig6}, we observe that the strength of instabilities are generally decreasing with increasing mass which corresponds to a decreasing luminosity-to-mass ratio.

\begin{figure*}
    \centering
    \begin{subfigure}{0.45\textwidth}
        \centering
        \includegraphics[width=\textwidth]{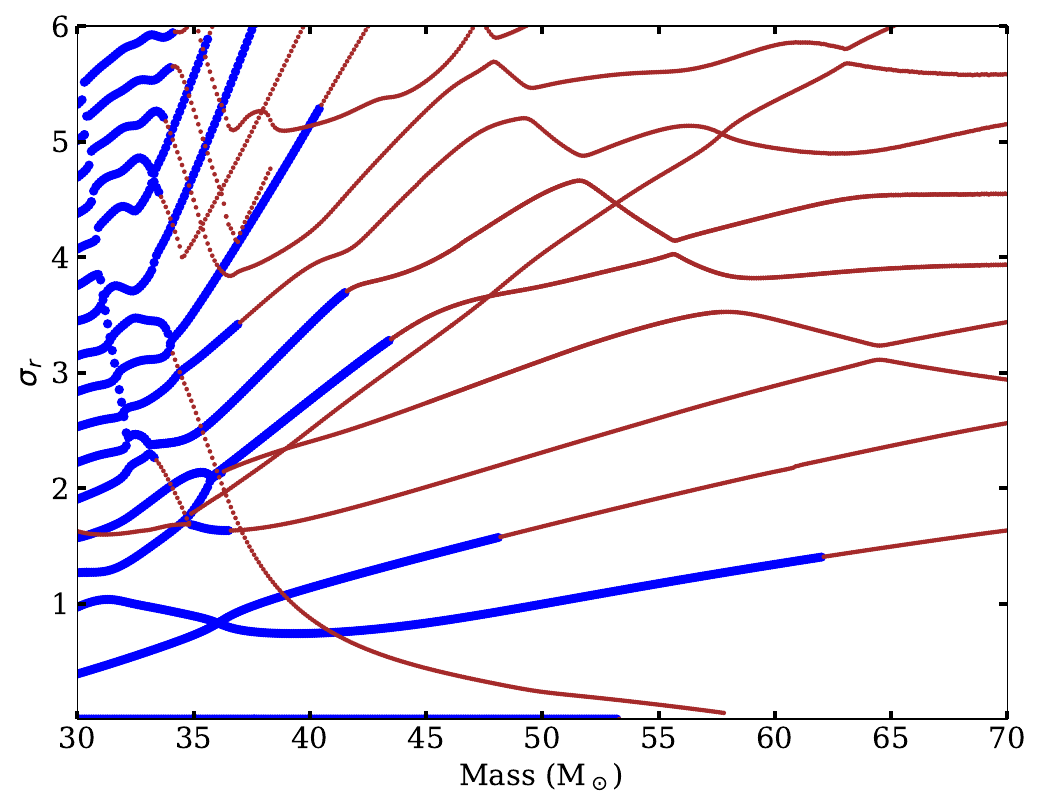}
        \label{fig8:sub1}
    \end{subfigure}
    \hfill
    \begin{subfigure}{0.45\textwidth}
        \centering
        \includegraphics[width=\textwidth]{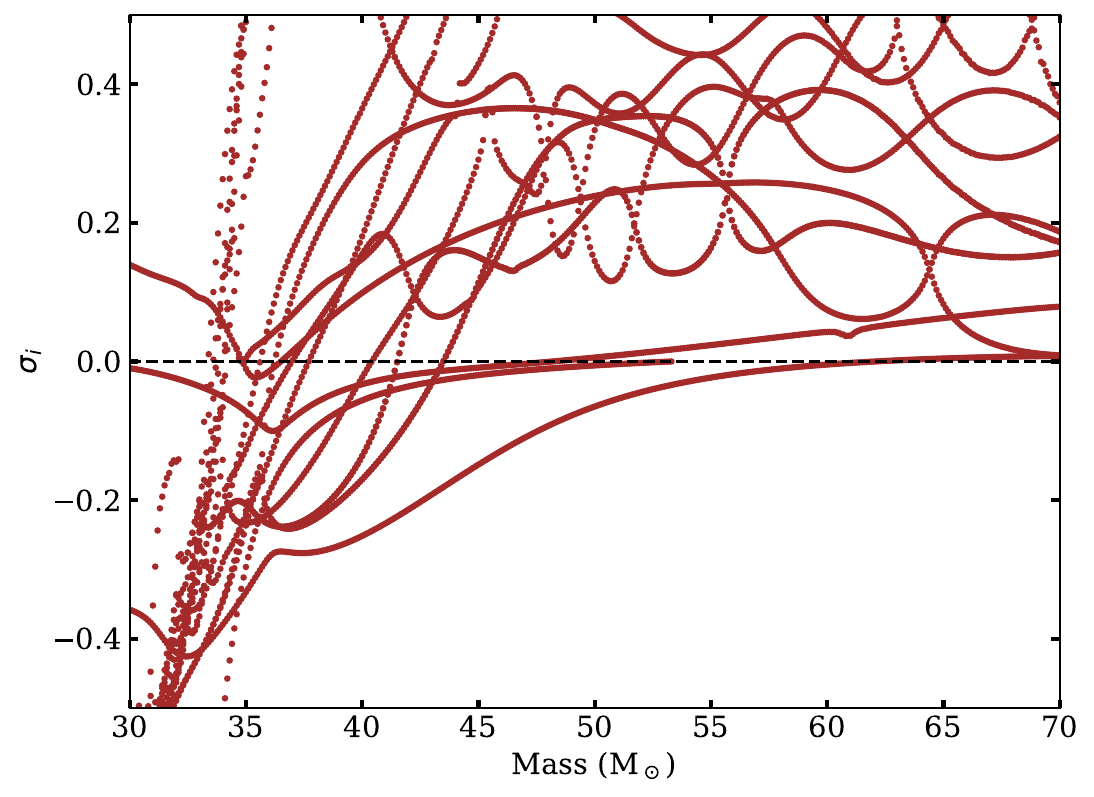}
        \label{fig8:sub2}
    \end{subfigure}
    \caption{Same as Fig.~\ref{fig6} but with boundary conditions consistent with those used in subsequent non-linear simulations.}
    \label{fig8}
\end{figure*}

\begin{figure*}
    \centering
    \begin{subfigure}{0.45\textwidth}
        \centering
        \includegraphics[width=\textwidth]{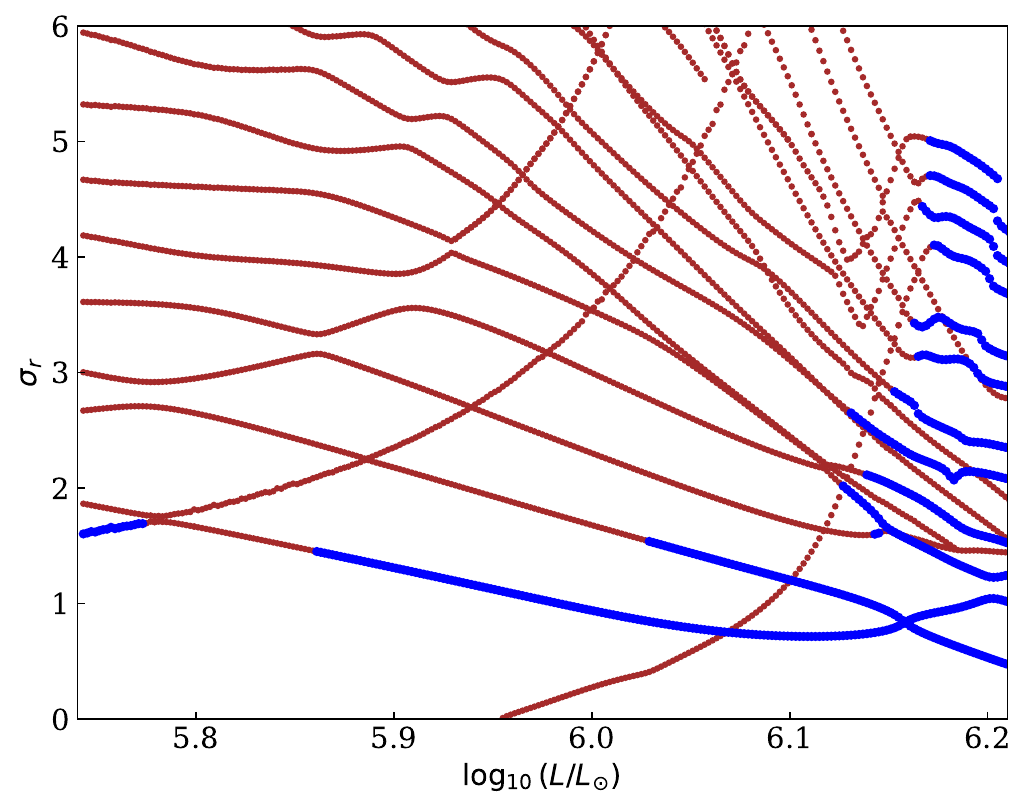}
        \label{fig9:sub1}
    \end{subfigure}
    \hfill
    \begin{subfigure}{0.45\textwidth}
        \centering
        \includegraphics[width=\textwidth]{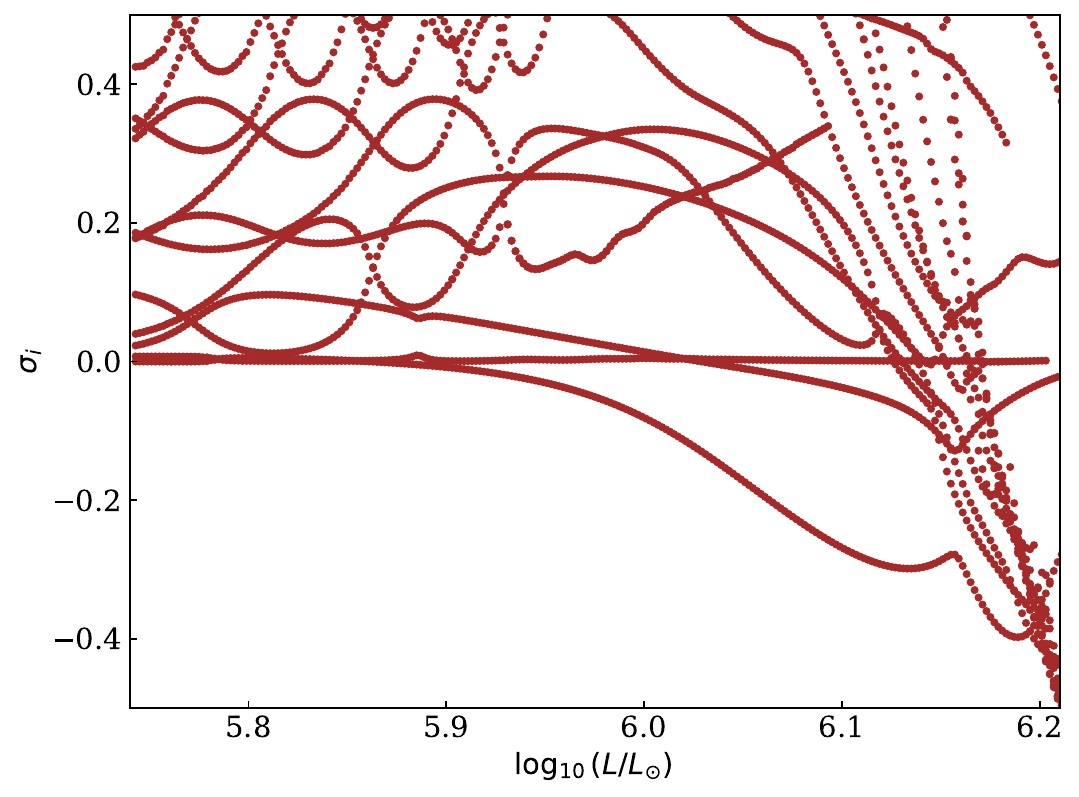}
        \label{fig9:sub2}
    \end{subfigure}
    \caption{Real (left) and imaginary(right) parts of eigenfrequencies plotted as a function of luminosity for models having a mass of 56.3 M$_{\odot}$. 
    The number of unstable modes increases with luminosity-to-mass ratio. }
    \label{fig9}
\end{figure*}

\begin{figure*}
    \centering
    \begin{subfigure}{0.45\textwidth}
        \centering
        \includegraphics[width=\textwidth]{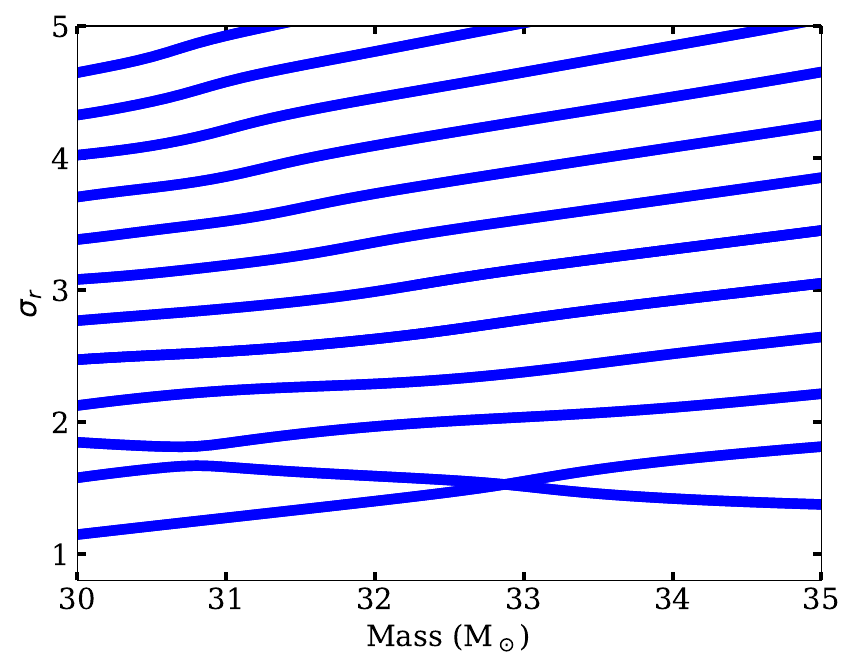}
        \label{fig11:sub1}
    \end{subfigure}
    \hfill
    \begin{subfigure}{0.45\textwidth}
        \centering
        \includegraphics[width=\textwidth]{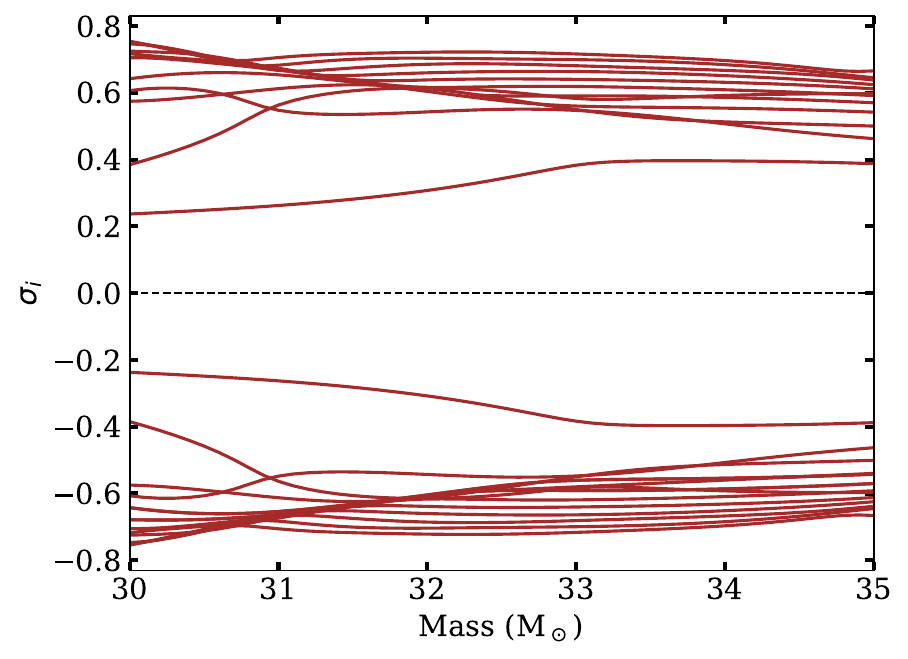}
        \label{fig11:sub2}
    \end{subfigure}
    \caption{Modes in the non-adiabatic reversible (NAR) approximation: Complex conjugate modes appear in the modal diagram.}
    \label{fig11}
\end{figure*}

Fig.~\ref{fig7} shows the pulsation periods associated with the modes as a function of mass. As before, the thick blue lines indicate unstable modes. Periods of the unstable modes range from less than 1 d to 7 d for models having mass below 35 M$_\odot$. Between 35 M$_\odot$ and 45 M$_\odot$, periods ranging from 1.5 d to 3.9 d are found. Beyond models of mass 45.6 M$_\odot$, the periods, corresponding to a single mode, lie between 1.6 to 3.2 d. 

There exists an ambiguity in the outer boundary conditions for the perturbation equations since the outer boundary of stellar model does not coincide with the physical boundary of the star. In some stellar models, changing the outer boundary conditions did not substantially change the results of the linear stability analysis \citep{yadav2021}. In other cases, the choice of outer boundary conditions was found to affect the instabilities in regions close to the stellar surface \citep{yadav2016, yadav2018}. Therefore, to check the dependence of the results of the linear stability analysis on the choice of outer boundary conditions, we repeat the linear stability analysis for the models considered here with different outer boundary conditions. The divergence of the heat flux and the gradient of the divergence of the velocity are required to vanish at the outer boundary in order to avoid the reflection of sound waves and shocks there \citep{grott2005}. These boundary conditions are also consistent with those used in the subsequent non-linear simulations. The results of the linear stability analysis with the altered boundary conditions is shown in Fig.~\ref{fig8}. A comparison with Fig.~\ref{fig6} reveals a similar behaviour in general. The monotonically unstable mode is present below 55 M$_\odot$. The two lowest order modes exhibit almost identical behavior, with the domain of the less strongly unstable mode increasing slightly to 47 M$_\odot$. However, the range of instability of some of the subsequent higher order excited modes seems to increase significantly. These modes, then may have significant amplitudes of the eigenfunctions at the outer boundary.   

There is a large uncertainty in the luminosity of $\epsilon$ Ori due to the uncertainties in the determination of its distance. The luminosity in \citet{puebla2016} is given as $\log(L_*/L_{\odot})$ = $5.92^{+0.32}_{-0.18}$.  In order to check the dependence of instabilities on the  luminosity, we perform a linear stability analysis for models having mass 56.3 M$_\odot$ and solar chemical composition, over a range of luminosities from $\log(L_*/L_{\odot})$ = 5.74 to $\log(L_*/L_{\odot})$ = 6.22. The results are given in Fig.~\ref{fig9}. Here, real and imaginary parts of the eigenfrequency are plotted as a function of luminosity. As the luminosity-to-mass ratio increases, the number of unstable modes as well the growth rates of the instabilities increases significantly. Above a luminosity of $6.1$, a large number of modes are unstable. Mode interaction is also frequent, which strongly indicates the presence of strange modes at high luminosities.

\subsection{NAR approximation}

\begin{figure}
    \centering
    \includegraphics[width=0.95\columnwidth]{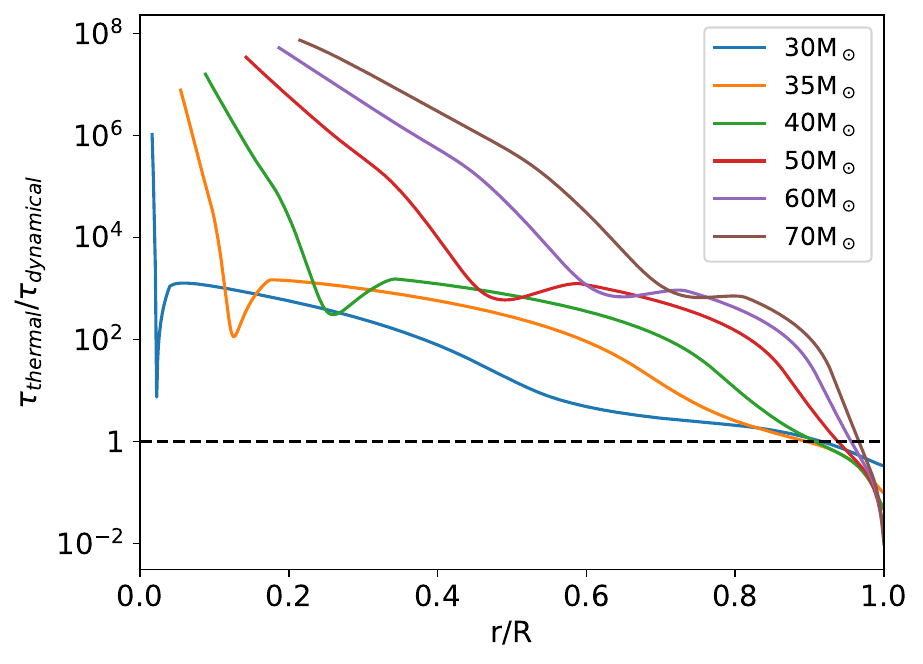}  
    \caption{Ratio of thermal to dynamical timescales as a function of relative radius for envelope models of different masses.}
    \label{fig10}
\end{figure}

The NAR (Non-Adiabatic-Reversible) approximation consists of setting the time derivative of the entropy in the perturbation equations to zero \citep{gg1990a}. As a result, no heat exchange between the different elements of the star are possible and 
any modes which are due to a heat driven instability (such as $\kappa$ and $\epsilon$ mechanism) are eliminated in this limit. Strange mode instabilities, however, are found to exist in the limit of the NAR approximation. This suggests that they are not related to the classical $\kappa$-mechanism. The NAR approximation thus provides a useful tool for classifying these modes and identifying the origin of the instabilities. 

The NAR approximation implies that the ratio of the local thermal timescale to the local dynamical timescale in the star \citep{gg1990a,glatzel2024} vanishes. Thus, an application of the NAR can be expected to provide at least qualitatively correct results only in stellar models having a low value of the local thermal timescale. The ratio of thermal to dynamical timescales as a function of relative radius for a few selected envelope models are shown in Fig.~\ref{fig10}. This ratio attains very high values in the stellar interior and decreases towards the surface. The region of the envelope over which the ratio of the thermal to dynamical timescale falls to order unity or below gradually decreases for higher masses. Thus, the NAR approximation seems to be justified only for the lower mass models of $\epsilon$ Ori, which also have a high ratio of luminosity-to-mass. In performing the NAR approximation on models of this star, we have therefore restricted ourselves to a mass range of $30-35$ M$_\odot$. 
Results of the NAR approximation are shown in Fig.~\ref{fig11}. The eigenvalues under the NAR approximation appear as complex conjugate pairs. Excited and damped modes may appear simultaneously, having identical values of the real part (frequency) and equal but opposite values of the growth and damping rates. Considering the dynamical instabilities at lower masses having very high growth rates, the NAR approximation seems to confirm the presence of strange modes in models below approximately 35 M$_\odot$ which have a luminosity-to-mass ratio greater than $2.4\times10^4\frac{L_\odot}{M_\odot}$.

\begin{figure*}
    \centering
    \begin{subfigure}{0.45\textwidth}
        \centering
        \includegraphics[width=\textwidth]{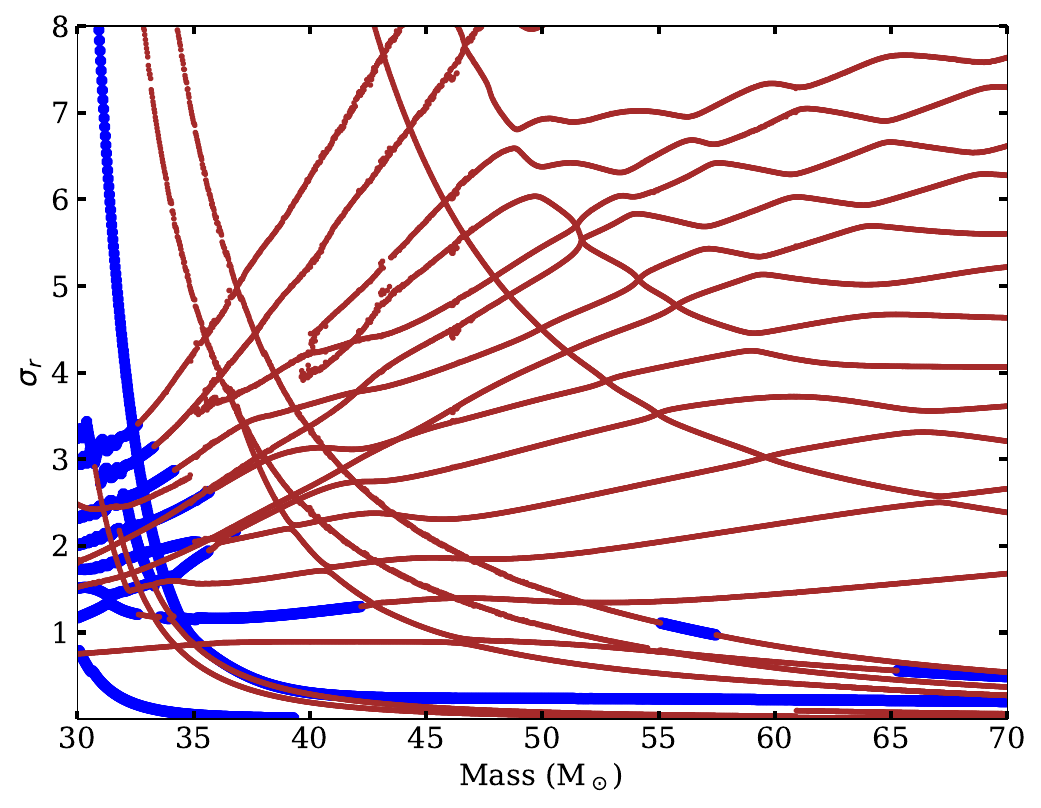}
        \label{fig12:sub1}
    \end{subfigure}
    \hfill
    \begin{subfigure}{0.45\textwidth}
        \centering
        \includegraphics[width=\textwidth]{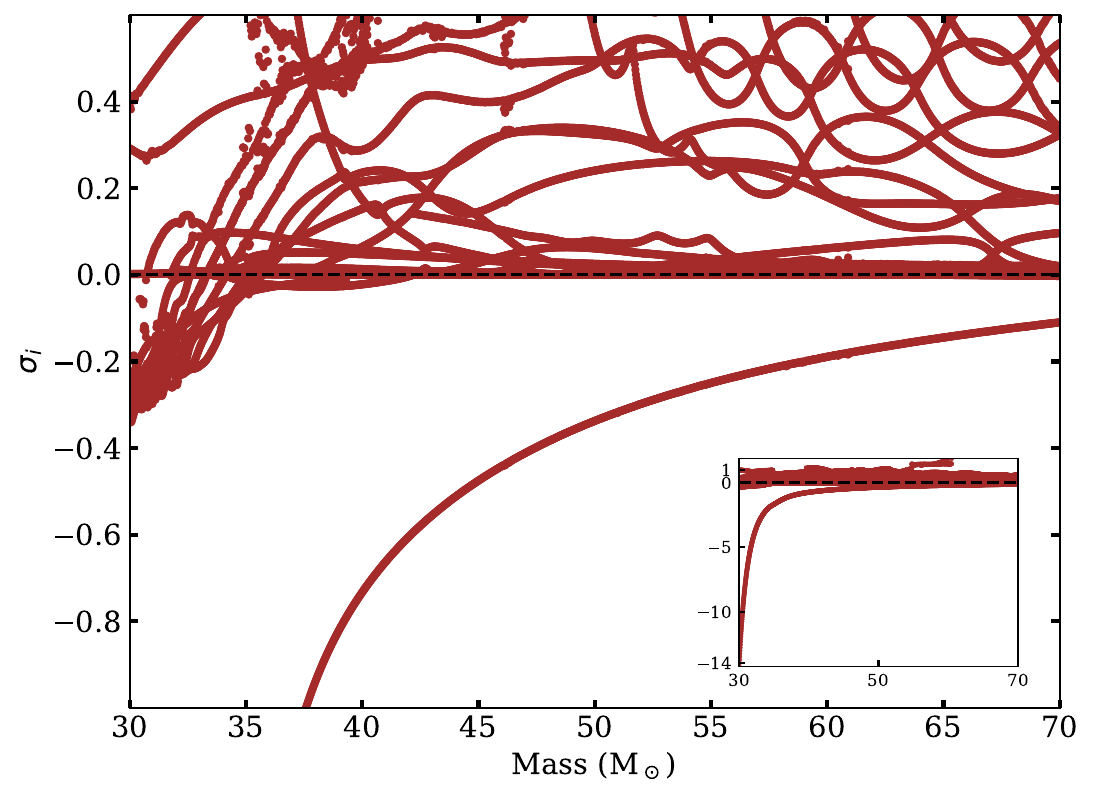}
        \label{fig12:sub2}
    \end{subfigure}
    \caption{Real (left) and imaginary (right) parts of eigenfrequencies plotted as a function of mass for harmonic degree $l$ = 2. Variation in  the imaginary part of the most unstable mode for lower mass models are given in the inset. }
    \label{fig12}
\end{figure*}

\subsection{Non-radial analysis}

\begin{figure*}
    \centering
    \begin{subfigure}{0.45\textwidth}
        \centering
        \includegraphics[width=\textwidth]{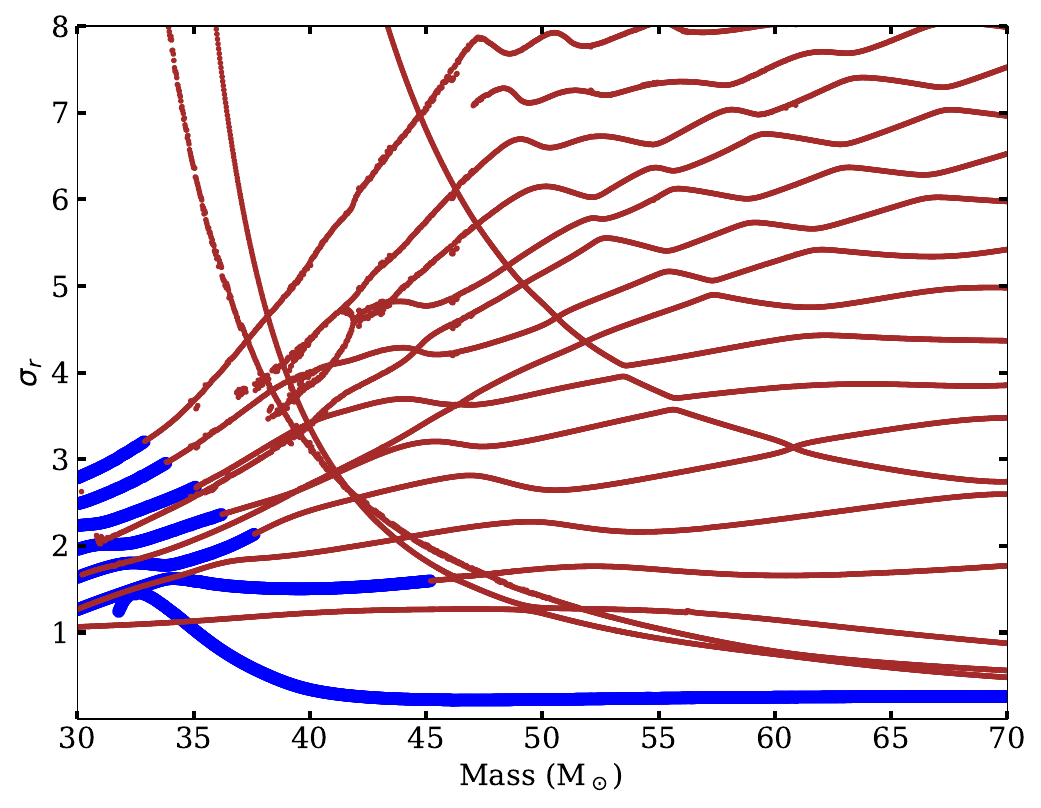}
        \label{fig13:sub1}
    \end{subfigure}
    \hfill
    \begin{subfigure}{0.45\textwidth}
        \centering
        \includegraphics[width=\textwidth]{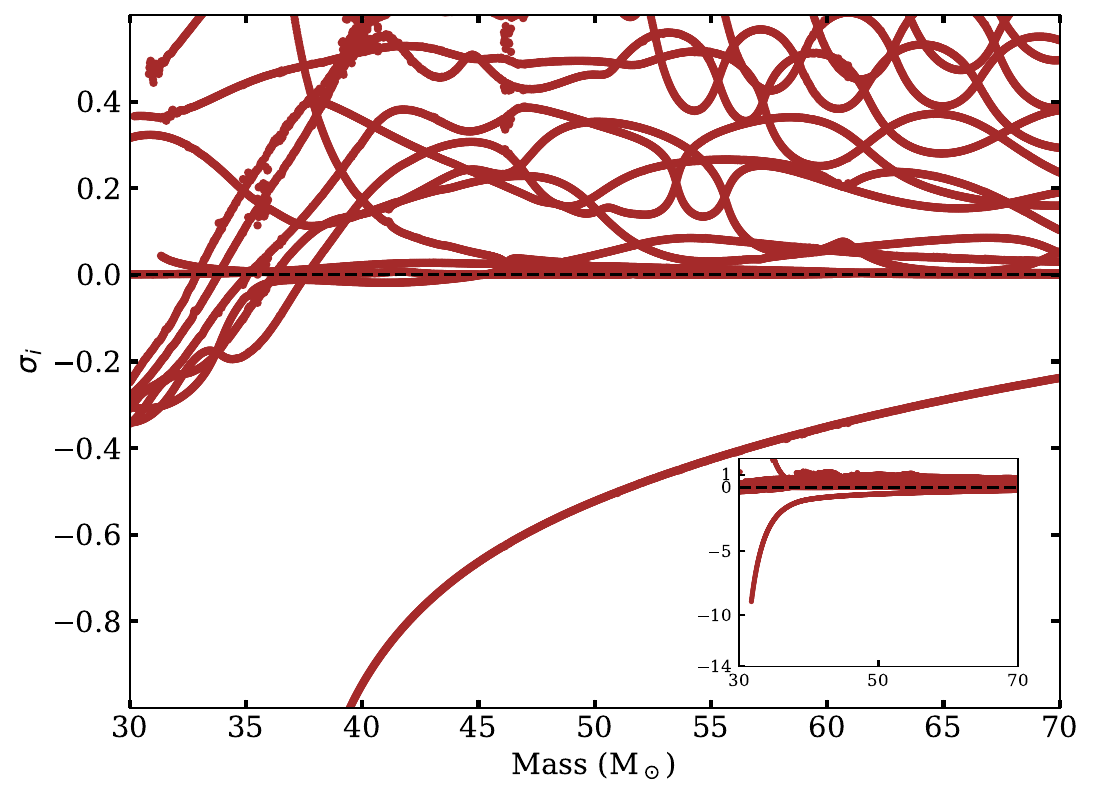}
        \label{fig13:sub2}
    \end{subfigure}
    \caption{Same as Fig.~\ref{fig12} but for harmonic degree $l$ = 4.}
    \label{fig13}
\end{figure*}

\begin{figure*}
    \centering
    \begin{subfigure}{0.45\textwidth}
        \centering
        \includegraphics[width=\textwidth]{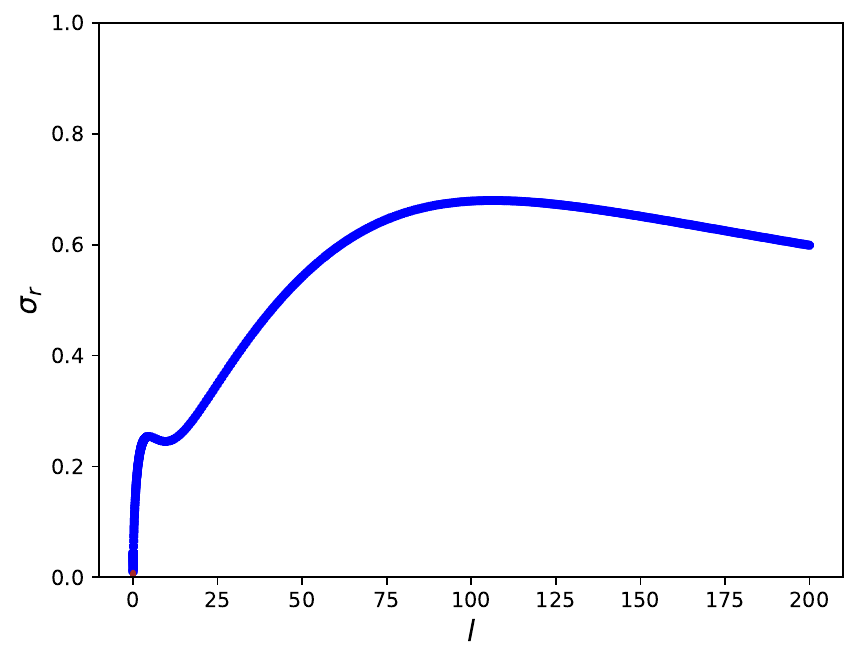}
        \label{fig14:sub1}
    \end{subfigure}
    \hfill
    \begin{subfigure}{0.45\textwidth}
        \centering
        \includegraphics[width=\textwidth]{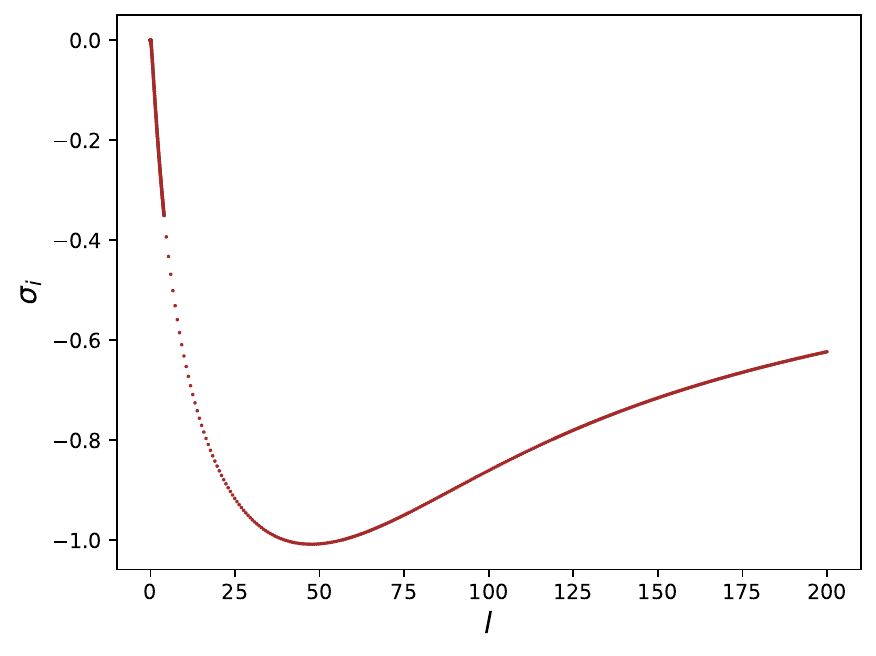}
        \label{fig14:sub2}
    \end{subfigure}
    \caption{Real (left) and imaginary (right) parts of the excited eigenfrequency for a 60 M$_\odot$ model as a function of harmonic degree $l$.}
    \label{fig14}
\end{figure*}

Massive stars, including supergiants, are known to exhibit non-radial pulsations \citep{saio2006, lefever2007, aerts2009}. In an earlier study by \cite{glatzel1996}, presence of dynamical mode-coupling instabilities have been identified in models of massive stars for non-radial perturbations.
Thus, a linear stability analysis with respect to non-radial perturbations seems to be appropriate for the models considered here. 
A non-radial stability analysis is performed for the same set of models referred to in Section \ref{models}. The equations describing the non-radial non-adiabatic pulsations are adopted in the form derived by \citet{glatzel1992}.This system of equations consists of the linearized forms of mass, momentum and energy conservation, Poisson’s equation for the gravitational potential, the energy transport equation and an equation of state. Angular dependence of the perturbations can be separated by an expansion in terms of spherical harmonics. Introducing an exponential time dependence of the perturbations, the equations are finally reduced to a sixth order system of ordinary differential equations. In order to solve this system of equations, six boundary conditions are required. The outer boundary conditions are ambiguous as the outer boundary of the model may not correspond to the physical outer boundary of the star. As in the case of the radial problem, the Stefan-Boltzmann law is considered to hold at the photosphere and the Lagrange pressure perturbation is assumed to vanish there. In addition, we require that at $x=1$ (where $x$ denotes the relative radius), the gravitational potential and its derivative are continuously connected to the vacuum solution for $x>1$.  The resultant sixth-order boundary eigenvalue problem is also solved using the Riccati method \citep{gg1990a}, giving us a set of complex eigenfrequencies.

The results are shown, as in the radial case, in the form of modal diagrams, where the real and imaginary parts of the eigenfrequencies are plotted as a function of mass for a fixed harmonic degree $l$. Outcomes of the non-radial non-adiabatic analysis are presented in Fig.~\ref{fig12} and Fig.~\ref{fig13} for harmonic degrees $l = 2$ and $l = 4$, respectively. Similar to the radial linear stability analysis, several modes are excited in models with higher luminosity-to-mass ratios. For both of the considered harmonic degrees ($l = 2, 4$), one of the modes is strongly unstable (see Figs. \ref{fig12} and \ref{fig13}). The real part of this highly unstable mode is almost constant for models having masses greater than $40\,M_{\odot}$. However, for less massive models, the mode is very sensitive to the stellar parameters. The strength of the instabilities associated with this mode is extremely high (of the order of the dynamical timescale) in models having mass close to $30\,M_{\odot}$, as depicted in the inset of imaginary part of the modal diagram. Based on their sensitivity to the stellar parameters, the modes in the modal diagram for $l = 2$ and $l = 4$ can be grouped into two categories. One set of modes exhibits higher frequencies in lower-mass models and lower frequencies in higher-mass models. The other set of modes shows the opposite behavior for models having mass below $47\,M_{\odot}$. To find the presence of instabilities for higher harmonics, we have selected the highly unstable mode for the $60\,M_{\odot}$ model. The variation of the real and imaginary parts of the eigenfrequency as a function of the harmonic degree up to $l$ = 200 is mentioned in Fig.~\ref{fig14}. The considered mode is unstable up to $l$ = 200 and appears to be unstable even for higher harmonics beyond $l$ = 200. The strength of instability is maximum around $l$ = 48 and decreases thereafter.


\section{Non-linear simulations} \label{nonlinear}

Several radial modes are excited in models of $\epsilon$ Ori. Qualitatively, the excited modes are not affected by the choice of our adopted boundary conditions. In order to find out the final fate of unstable models, we have adopted a non-linear numerical scheme proposed by \citet{grott2005}.  
Constant luminosity and zero velocity amplitude are used as the bottom boundary conditions. Vanishing gradient of the divergence of velocity and the divergence of the heat flux are adopted for surface boundary conditions. The outer boundary conditions are chosen in such a way that the outgoing shock waves can pass the boundary without reflection. 
Suitability of the adopted numerical scheme has been presented in earlier studies \citep[see e.g.,][]{yadav2017, yadav2021}. In earlier studies \citep{yadav2017zams}, it has been shown that the code picks up the instability from the numerical noise level of the order of 10$^{-5}$ cm/s without adding any external perturbation. In the linear phase of the exponential growth, the period and growth rate obtained from the simulation are identical to those obtained from the previous independent linear analysis.

\begin{figure*}
    \centering
    \includegraphics[width=\textwidth]{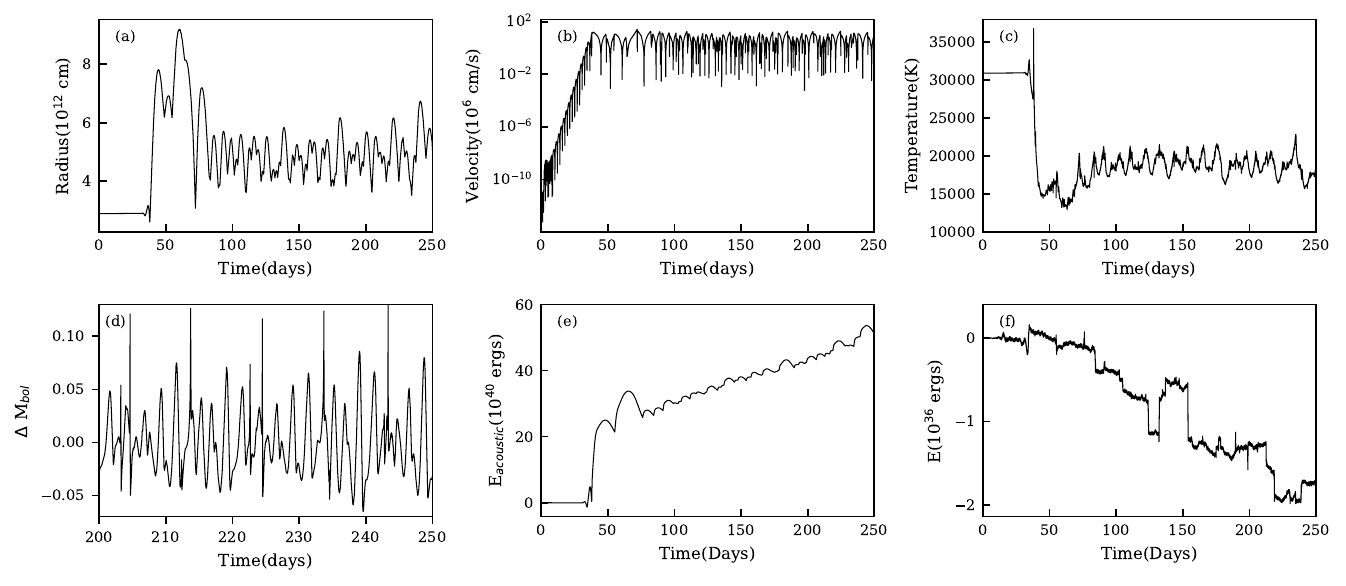}  
    \caption{The non-linear evolution of instability for a 41 M$_\odot$ model with solar chemical composition. The following quantities are shown as a function of time: (a) stellar radius, (b) velocity and (c) temperature at the photosphere and (d) variation of the bolometric magnitude , (e) time-integrated acoustic luminosity and (f) error in the energy balance. Finite amplitude pulsation without any strict periodicity is found to be the result of the instability. }
    \label{fig15}
\end{figure*}

\begin{figure*}
    \centering
    \includegraphics[width=\textwidth]{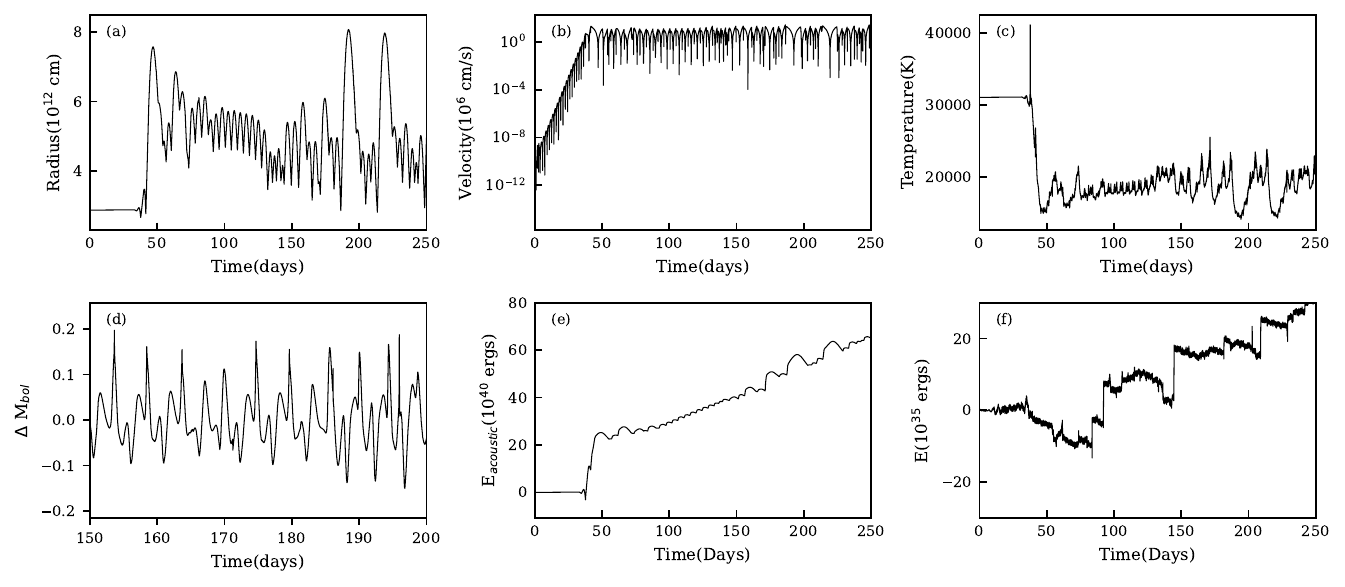}  
    \caption{Same as Fig.~\ref{fig15} but for a model of 45 M$_\odot$. As in the case of the 41 M$_\odot$ model, the bolometric magnitude variation shows finite amplitude pulsation but with no clear period. }
    \label{fig16}
\end{figure*}


\begin{figure*}
    \centering
    \includegraphics[width=\textwidth]{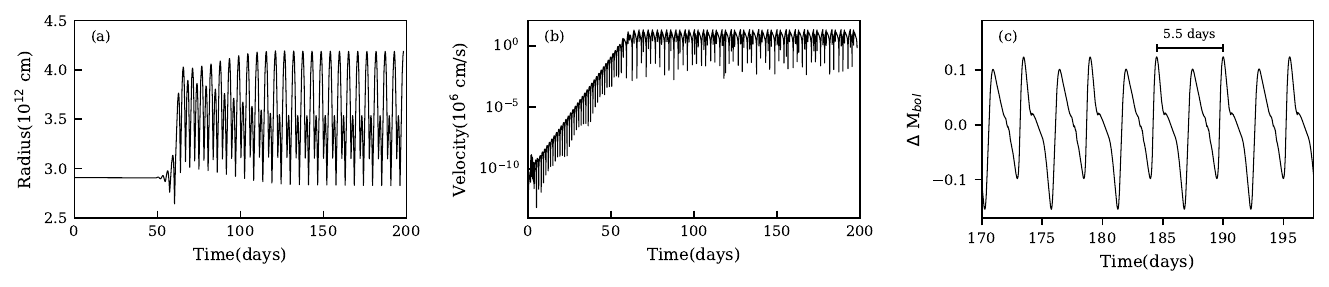}  
    \caption{Evolution of instability into the non-linear regime for a model of 50 M$_\odot$ . The radius (a), velocity (b) at the photosphere and the variations of the bolometric magnitude (c) as a function of time are shown. In the linear phase of exponential growth, the period of 2.8 d is consistent with the period obtained from the independent linear analysis. In the non-linear regime, instability leads to finite amplitude pulsation with an increased period of 5.5 d. This is due to the inflation of the stellar envelope.}
    \label{fig17}
\end{figure*}

\begin{figure*}
    \centering
    \includegraphics[width=\textwidth]{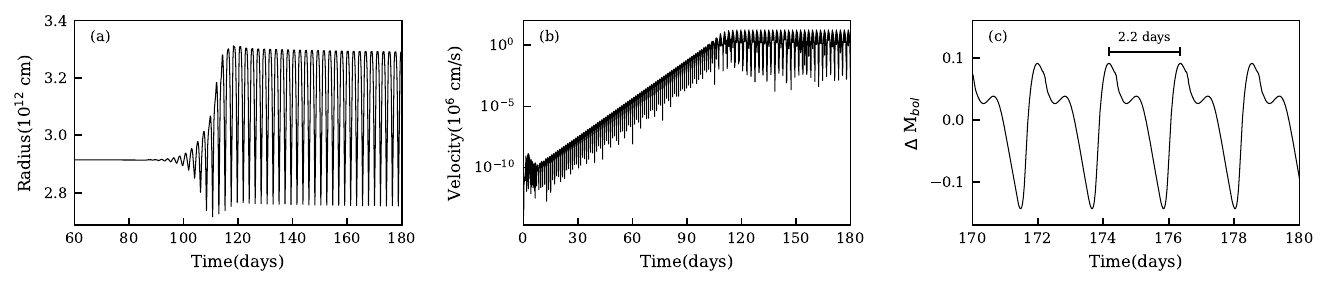}  
    \caption{Same as Fig.~\ref{fig16} but for a model with mass 56.3 M$_\odot$. The adopted mass of the model is consistent with mass of $\epsilon$ Ori as found from its evolutionary tracks. Here the instability leads to finite amplitude pulsation with a period of 2.2 d.  }
    \label{fig19}
\end{figure*}

In carrying out simulations of the instabilities and pulsations in the non-linear regime, a correct treatment of the energy balance is essential. The thermal and gravitational energies of the system exceed the kinetic energy by at least two orders of magnitude; the acoustic luminosity being at least three orders of magnitude smaller than the maximum energies. Thus, a correct determination of the acoustic luminosity requires the relative error in the energy balance to be less than 10$^{-4}$. Such a high accuracy is obtained by the use of a fully conservative scheme as described in \citet{grott2005} which ensures the reliability of the outcomes of the non-linear simulation.

\subsection{Results of the non-linear simulations}

We have performed non-linear simulations for selected stellar models in the mass range of 30 to 70 M$_\odot$. 
The considered models have effective temperature $T_\mathrm{eff}$ = 27000 K and  luminosity $\log (L/L_\odot)$ = 5.92 as adopted in the linear stability analysis.
The results of non-linear simulation for a model of 41 M$_\odot$ are given in Fig.~\ref{fig15} , which shows the radius (a), velocity (b), temperature (c) and bolometric magnitude (d) at the photosphere together with the time integrated acoustic energy (e) and the error in the energy balance (f) as a function of time. The simulation starts from the level of numerical noise and without any external perturbation picks up the instability. It then exhibits a linear phase of exponential growth and the velocity amplitude finally saturates in the non-linear regime after about 100 days. The radius is significantly inflated initially before showing irregular variations at a comparatively lower value. The variation of the bolometric magnitude shows finite amplitude pulsation without any well-defined period. Such behavior has been reported in non-linear simulations of massive OB type stars \citep{yadav2017} and is consistent with the observed irregular variability in the $\alpha$ Cygni variables. 
The acoustic energy lost by the system is found to increase in a non-monotonic manner with time. This is because during a cycle of pulsation, both incoming and outgoing acoustic fluxes exist; however the outgoing flux exceeds the incoming flux and hence the average over one pulsation cycle of the time integral of the acoustic luminosity should increase with time as we see in this case. The slope of the curve thus corresponds to a mean acoustic luminosity driven by pulsation.
The error in the energy balance (Fig.~\ref{fig15}f) is smaller than the acoustic energy (the smallest term in the energy balance) by at least five orders of magnitude. The results of the non-linear simulation for a 45 M$_\odot$ are shown in Fig.~\ref{fig16}.  The radius is considerably inflated in the non-linear regime due to successive shock waves. Similar to the case of the model with 41 M$_\odot$, the bolometric magnitude shows a finite amplitude pulsation with no strict periodicity.

In Fig.~\ref{fig17}, the non-linear simulation of a model of 50 M$_\odot$ is shown. Linear analysis reveals an unstable mode with $\sigma_r$ = 0.98 corresponding to a period of 2.80 days. In the linear phase of the numerical simulation, the period obtained is 2.89 days, which is close to that derived from the independent linear stability analysis. In the non-linear regime, the velocity amplitude saturates at 137 km s$^{-1}$. The variation of the bolometric magnitude shows a periodic finite amplitude pulsation with a period of 5.5 days. The increase in the period is due to the inflation of the radius in the non-linear domain as compared to the initial hydrostatic value. Previous non-linear studies \citep{glatzel2009, yadav2016} have shown that strong instabilities may considerably inflate the stellar envelope due to which the final non-linear pulsation periods differ significantly from the linearly determined periods. As such, the observed periods should be compared with the periods obtained from the non-linear simulation rather than the linear periods. 
For the non-linear simulation of a 55 M$_\odot$ model, the instability again leads to a finite amplitude pulsation of 2.4 days and the velocity amplitude saturates at 123 kms$^{-1}$. 

The location of $\epsilon$ Ori on the HR diagram indicates that model with parameters close to the star has a mass of 56.3 M$_\odot$ (see Fig.~\ref{fig5}). This mass is likely to be an upper limit as the inclusion of rotation corresponds to a lower value of mass for the given luminosity and effective temperature on evolutionary tracks.  
Therefore we have also considered a model with this mass for the non-linear simulation which has been presented in Fig.~\ref{fig19}. 
 After the linear phase of exponential growth the velocity amplitude in the non-linear regime saturates at 131 km s$^{-1}$. 
 The simulation finally converges to a finite-amplitude pulsation with a period of 2.2 days, as indicated by the bolometric magnitude profile (Fig.~\ref{fig19}c). Inflation in the radius is also noticeable as found in the earlier mentioned models. For the model of 60 M$_\odot$, velocity amplitude saturates at 135 km s$^{-1}$ and the bolometric magnitude shows finite amplitude pulsation with a period of 1.9 days.

\begin{figure*}
    \centering
    \includegraphics[width=0.95\textwidth]{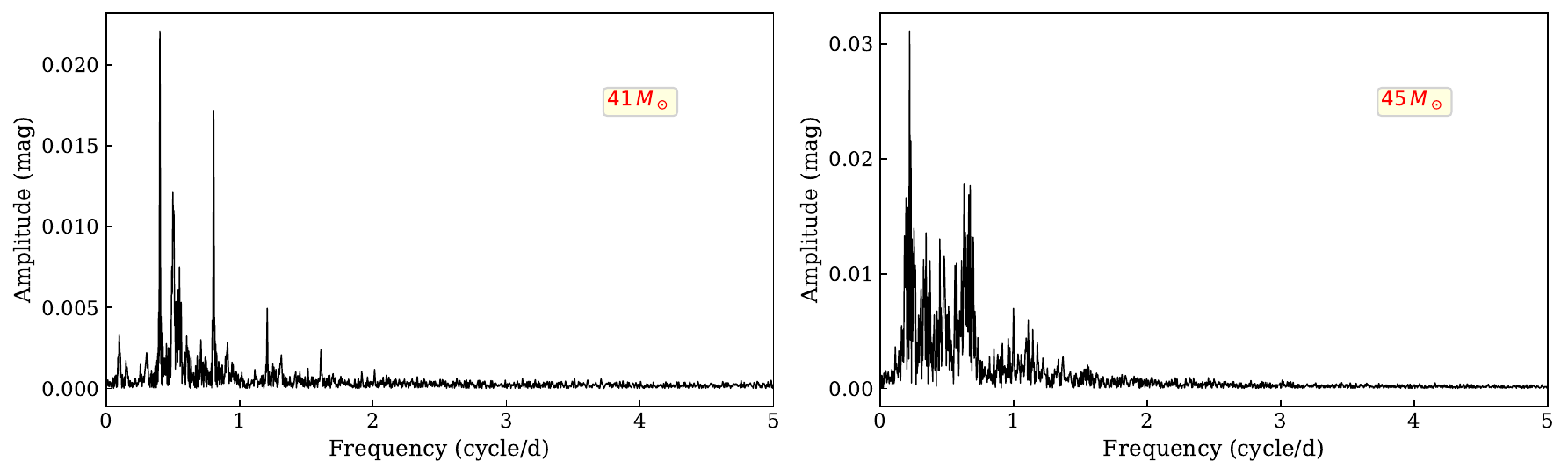}
    \caption{Periodograms obtained from the variation of the bolometric magnitude at the photosphere for the 41 M$_\odot$ (left) and 45 M$_\odot$ model (right). The variation of the bolometric magnitude for computing the periodograms is taken after the amplitude saturation from 100 d to 250 d. The dominant frequency for the periodogram for 41 M$_\odot$ is 0.40 c/d while that for the 45 M$_\odot$ is 0.22 c/d. }
    \label{fig:period_41_45}
\end{figure*}

The lower mass models are showing irregular variations in the non-linear regime (see Figs. \ref{fig15} and \ref{fig16}), which have also been found in $\alpha$ Cygni variables. We have obtained the periodograms for the variations in the bolometric magnitude of two models having a mass of 41 M$_\odot$ and 45 M$_\odot$ using PERIOD04 \citep{2004IAUS..224..786L}. The outcome is shown in Fig. \ref{fig:period_41_45} for these two models. Similar to the observed TESS periodograms (Fig.~\ref{fig1}), the frequency content of the theoretical periodograms lie below 2 d$^{-1}$. The theoretical periodograms show coherent peaks and the dominant frequencies for the 41 M$_\odot$ and 45 M$_\odot$ are found to be 0.40 d$^{-1}$   and 0.22 d${^{-1}}$ respectively. The latter is of the order of the marginally dominant frequency of 0.245 d$^{-1}$ obtained from the TESS periodogram of Sector 6.   
The theoretical periodograms consist of only radial modes, while the observed periodograms are expected to have both radial and non-radial modes. Therefore, it is difficult to directly compare these periodograms.  However, the order of the dominant frequencies obtained from the theoretical periodograms is also consistent with earlier observed spectroscopic periods.

\begin{figure*}
    \centering
    \includegraphics[width=\textwidth]{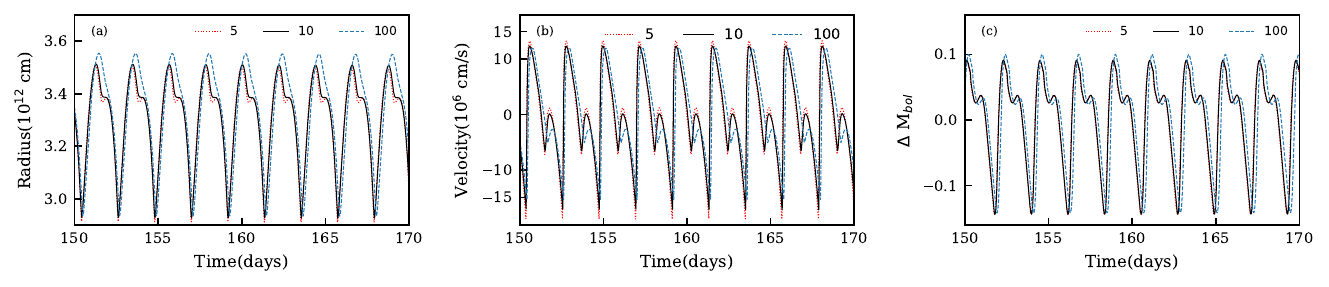}  
    \caption{The influence of the artificial viscosity parameter ($\nu_0$ = 5,10,100) on the finite amplitude pulsations of a 56.3 M$_\odot$ model. The radius (a), velocity (b) and variation of the bolometric magnitude (c) are given at the outermost grid point as a function of time.}
    \label{fig21}
\end{figure*}

\subsection{Effect of artificial viscosity in the non-linear regime}

Strong shock waves are expected to occur when the instabilities are followed into the non-linear regime. The discontinuities caused by these shocks usually have to be spread out over a number of mass zones in order to avoid the very large gradients that are associated with this phenomenon. One method to do this is to introduce a so-called artificial viscosity into the system. Although there are many forms of artificial viscosity, we use the one considered by \citet{grott2005}. The quality and reliability of the results depend on a correct value of the artificial viscosity. A very high value of the artificial viscosity would damp the physical instabilities. If it is too small, numerical oscillations occur behind the shock front (Gibbs phenomena). To avoid such numerical artifacts, the artificial viscosity has to be chosen carefully. As such, the effect of the artificial viscosity, in particular the artificial viscosity parameter $\nu_0$ \citep{grott2005} on the outcomes of the non-linear simulations need to be studied.

In general, the viscosity parameter is to be chosen as small as possible but sufficiently high in order to suppress the Gibbs phenomena. A value of $\nu_0$=10 has been found to be appropriate in many previous studies \citep{grott2005,yadav2016,yadav2017}. The effect of $\nu_0$ on the non-linear evolution of the 56.3 M$_\odot$ model is presented in Fig.~\ref{fig21} for three values of $\nu_0$: 5, 10 and 100. It appears that the radius, velocity and variation of bolometric magnitude do not show much significant change when the artificial viscosity is varied.

\section{Discussion} \label{discussion}

$\alpha$~Cygni variables exhibit both photometric and spectroscopic variability on various timescales. In this study, we analyzed the available \textit{TESS} data for the $\alpha$ Cygni variable HD~37128 ($\epsilon$~Ori). The star was observed in two \textit{TESS} sectors, and was found to show only marginally dominant peaks at frequencies of 0.24 and 0.74~c/d in Sectors~6 and~32, respectively. Our analysis also reveals the presence of stochastic low-frequency variability in both sectors, consistent with previous findings by \citet{bowman2020}. Although the TESS photometry is dominated by stochastic variability, the pulsation analysis presented is motivated by the periodicities reported from spectroscopic studies. In earlier spectroscopic analyses of HD 37128, variability of the order of 2 to 8 days has been reported \citep[see e.g.,][]{morel2004, prinja2004,thompson2013, martins2015}. 

To understand the pulsational instabilities and their connection with the observed variability in HD 37128, we have performed non-adiabatic radial and non-radial linear stability analyses in models of this star having a mass range between 30 and 70 M$_\odot$. The outcomes of the linear stability analysis have been presented in Section \ref{lna} . Frequent mode interaction has been noted in the modal diagrams for the radial as well as the non-radial analysis. The luminosity-to-mass ratio of the considered models is in the range of 3$\times$10$^4$ to 1$\times$10$^4$ in units of the solar luminosity-to-mass. The presence of strange mode instabilities is expected in these models as reported in earlier studies \citep{saio2013, yadav2017}. The period associated with the excited modes in the radial analysis is in the range of few hours to 6.8 days. These periods are consistent with those observed in this star. 
In order to understand the impact of boundary conditions on the outcome of our numerical results, we have performed the radial stability analysis with different boundary conditions \citep[see][]{grott2005}. Since the luminosity of the star is very uncertain, we have carried out a stability analysis for a given value of mass and effective temperature with the luminosity as a parameter. The number and strength of instabilities increase with increasing luminosity-to-mass ratio, consistent with earlier findings from models of massive stars \citep{yadav2017,parida2024,glatzel2024}. The models of $\epsilon$~Ori show pronounced core-envelope structure where the ratio of local thermal and dynamical timescales (see Fig.~\ref{fig10}) is considerably small in the envelope, satisfying key physical conditions for the occurrence of strange mode instability as mentioned by \citet{glatzel_1994}. The NAR approximation for low mass models has confirmed the presence of strange mode instabilities in the present study.
We have also performed stability analysis with respect to non-radial perturbations for the considered models. Instabilities are present in all the considered models in the adopted mass range for low harmonic degrees. Notably, instabilities are also found to be consistently present in high harmonic degrees, as noticed in models of Wolf-Rayet stars \citep[see e.g.][]{glatzel_2002}.

Non-linear simulations has been carried out in selected models of $\epsilon$~Ori to find out the consequences of identified instabilities. Earlier studies \citep[e.g.,][]{yadav2016, yadav2017, yadav2021} have shown that such instabilities can lead to finite amplitude pulsations, extreme envelope inflation, surface eruption and rearrangements of stellar structure. For the reliable outcomes of the non-linear simulations, we have ensured that the error in the energy balance is several order of magnitudes smaller than the energies involved. The importance of the energy balance has been highlighted by \citet{grott2005} and \citet{yadav2016}. In all the considered models, finite amplitude pulsation appears to be the final result of the instabilities. Notably, for the 41 M$_\odot$ and 45 M$_\odot$ model, where the growth rates are high ($\sigma_i<-0.1$), the pulsation does not exhibit a strictly periodic pattern. This is consistent with the luminosity variations observed in $\alpha$ Cygni stars \citep[see e.g.,][]{van1998}. While comparing with TESS observations, it should be noted that the variations of bolometric magnitudes are obtained from non-linear simulations, whereas TESS measures the flux variations in a broad optical bandpass. In addition, the non-linear simulations presented here are restricted to radial instabilities. The observed variability for such massive stars, however, is very likely to be a combination of radial and non-radial modes. Moreover, B-type supergiants such as $\epsilon$ Ori have strong stellar winds which vary with time (as seen from the H$\alpha$ profiles) and thus could introduce additional variability signatures on the observed photometric magnitudes for $\alpha$ Cygni variables \citep[see e.g.,][]{kraus2021}. Therefore, a direct comparison of the non-linear bolometric magnitude variations with the TESS data should be taken with caution. While the TESS lightcurves show an irregular behaviour comparable to the magnitude variations seen in the 41 M$_\odot$ and 45 M$_\odot$ models, the observed variations are of the order of $\sim$ 0.05 mag, which is less than that found theoretically. Further, the TESS periodograms do not show the coherent peaks which are found in the theoretical periodograms. Instead, they are contaminated with a significant amount of stochastic low-frequency variability (red noise, see Fig. \ref{fig2}).

For higher masses, periodic pulsation having periods 1.9 d to 6 d are found in the models. These periods lie well within the observed range of spectroscopic periods which have been found for this star \citep{prinja2004, thompson2013}. Particularly, for the 56.3 M$_\odot$ model, the final pulsation period (2.2 d) is close to the recurrent period found in the radial velocity data for several observing seasons \citep{thompson2013}. 
Also, for the 60 M$_\odot$ model, the pulsation period of 1.9 d agrees remarkably well with the most prominent period of 1.92 d found by \citet{prinja2004} in the wind and photospheric lines of $\epsilon$ Ori.  
Similar for the case of standard O and B type stars \citep{yadav2017}, the choice of artificial viscosity parameter does not considerably affect the pulsation period, velocity amplitude, and variation in the radius profile (see Fig.~\ref{fig21}). 

The variability of $\epsilon$ Ori is rather complex. As in the $\alpha$ Cygni variables, the multiperiodic variability may be a result of the simultaneous excitation of radial and non-radial instabilities \citep[see e.g.,][]{saio2011, saio2013}. However, these studies were restricted to a linear analysis only. For the strong instabilities (strange modes) which are excited in the models of these luminous stars, a non-linear analysis is required in order to compare theoretical predictions with observations. In the present study, both radial and non-radial instabilities have been found for models of this star. The radial instabilities when followed into the non-linear regime lead to finite amplitude pulsation. Strong non-radial instabilities were also found; however, following them into the non-linear regime is not yet feasible. Therefore, to understand what the final consequences of the radial and non-radial modes or their combinations can be in the star, the numerical representation of non-linear non-radial pulsations is necessary. Only then can a more realistic understanding of the pulsational instabilities and their connection with the observed variability of $\alpha$ Cygni stars be obtained.

Luminosity-to-mass ratio of some models of $\epsilon$ Ori indicates highly non-adiabatic conditions in the star. In extremely non-adiabatic conditions, due to frequent mode interactions, following the eigenfrequencies as a function of stellar parameter in the modal diagram becomes a challenging task. The Riccati method used to solve the pulsation equations by \citet{gg1990b} has been successfully applied in calculating eigenfrequencies for highly non-adiabatic and frequent mode interaction conditions \citep{gg1990a, glatzel1993}. Contrary to the weaknesses of the method mentioned by \citet{goldstein2020}, local minima in the determinant function along real as well as imaginary axes are used to find the approximate location of roots in extremely non-adiabatic conditions. A scan along the imaginary axis helps us to identify modes that are also far from the real axis. It also helps us identify monotonically unstable or stable modes \citep{yadav2016, yadav2018}. In addition to that, it is worth emphasizing that the method can successfully follow modes showing complex conjugate behavior, which is common in extremely non-adiabatic environments \citep[see e.g.,][]{gg1990a, kiriakidis1996, glatzel2024}.   

\section{Conclusions} \label{conclusions}

The findings of the present work are summarized below:
 
\begin{enumerate}
    \item In addition to the overall white noise, stochastic low-frequency variability is present in both TESS sectors for $\epsilon$ Ori. No significant peaks are detected in either sector, confirming previous results. The sector-to-sector changes in the SLF parameters, despite the identical cadence, are particularly noteworthy. Additional sector data will be necessary to better understand the origin and significance of these variations.

    \item In considered models of this star, radial and non-radial modes are unstable having periods in the range of a few hours to 12 days. The number and strength of instabilities are increasing for models having higher luminosity-to-mass ratios. Instabilities are also found to be present for high harmonic degrees ($l$ > 5).   

    \item Thermal and dynamical timescales become comparable in the outer parts of the models. The non-adiabatic reversible (NAR) approximation has been used to show the presence of strange mode instabilities with a non-thermal origin.  

    \item Numerical simulations in the non-linear regime have shown finite amplitude pulsation in models of $\epsilon$ Ori. The average radius of all the finite amplitude pulsating models has considerably increased compared to the initial hydrostatic value. 
    Models with a mass less than 50 M$_{\odot}$ exhibit semi-regular variability on a timescale consistent with observations. For higher-mass models, finite amplitude pulsation occurs with well-defined periods. While such behaviour is not typical for $\alpha$ Cygni variables, it is interesting to note that the periods agree well with some of the prominent spectroscopic periods found in previous studies. Thus, radial pulsations may contribute to some of the observed variability.   
\end{enumerate}

Long-term photometric and spectroscopic monitoring is required to determine variability on different time scales and variations in the wind and mass-loss rate of $\epsilon$ Ori. A comparison of the pulsation-driven mass-loss rate with the observed mass-loss rate will help us to strengthen our understanding of the pulsation and mass-loss interaction in $\alpha$ Cygni variables.

\section*{Acknowledgments}
SP acknowledges financial support from the DST–INSPIRE Fellowship (IF220167). MK acknowledges financial support from the Czech Science Foundation (GA\v{C}R, grant number 25-17532S). The Astronomical Institute of the Czech Academy of Sciences is supported by the project RVO:67985815. The project is co-funded by the European Union (Project 101183150 - OCEANS).

\section*{Data Availability}
TESS data used in this work are available for download from the Barbara A. Mikulski Archive for Space Telescopes (MAST) at \href{https://mast.stsci.edu}{mast.stsci.edu}. The model data underlying this study are available from the corresponding author upon reasonable request.



\bibliographystyle{mnras}
\bibliography{example} 





\bsp	
\label{lastpage}
\end{document}